\documentclass[
prb,
reprint,
floatfix,
amsmath,amssymb,
showpacs,
superscriptaddress,
citeautoscript,
nonatbib
]{revtex4-1}

\usepackage{xr}
\makeatletter
\newcommand*{\addFileDependency}[1]{
	\typeout{(#1)}
	\@addtofilelist{#1}
	\IfFileExists{#1}{}{\typeout{No file #1.}}
}
\makeatother

\newcommand*{\myexternaldocument}[1]{
	\externaldocument[S-]{#1}
	\addFileDependency{#1.tex}
	\addFileDependency{#1.aux}
}

\myexternaldocument{SM}

\usepackage{amssymb}
\usepackage[breaklinks]{hyperref}
\usepackage[all]{hypcap}
\hypersetup{
	plainpages=false,
	unicode=false,                          
	pdfborder={0 0 0},
	pdftoolbar=true,                        
	pdfmenubar=true,                        
	pdffitwindow=false,                     
	pdfstartview={FitH},                    
	pdftitle={},                            
	pdfauthor={},                           
	pdfproducer={LNCMI-CNRS-UGA-UPS-INSA},  
	pdfkeywords={High} {Magnetic} {Fields}, 
	pdfnewwindow=true,                      
	linktoc=section,
	colorlinks=true,                        
	linkcolor=blue,                         
	citecolor=red,                          
	filecolor=magenta,                      
	urlcolor=blue                           
}
\usepackage{graphicx}
\usepackage{float}
\usepackage{amsmath}
\usepackage{bm}
\usepackage{color, soul}
\usepackage[normalem]{ulem}
\usepackage{siunitx}
\usepackage[version=4]{mhchem}
\usepackage{upgreek}
\usepackage[usenames,dvipsnames]{xcolor}
\usepackage{textcomp}
\usepackage{chemformula}
\usepackage[normalem]{ulem}
\usepackage[english,capitalize]{cleveref}
\usepackage{acronym}
\usepackage{multirow}
\usepackage[utf8]{inputenc}
\usepackage{microtype}
\usepackage[T1]{fontenc}
\usepackage{siunitx}
\usepackage{comment}

\newcommand{\cpb}{\ce{CsPbBr3}}
\newcommand{\csb}{\ce{CsSrBr3}}
\newcommand{\csbe}{\csb:Eu}

\newacro{HaP}{Halide perovskite}
\newacro{XRD}{X-ray diffraction}
\newacro{IR}{infrared}
\newacro{DOS}{density of states}
\newacro{LPE}{lone pair of electrons}
\newacro{MD}{molecular dynamics}
\newacro{COHP}{crystal-orbital Hamilton population}
\newacro{VBM}{valence band maximum}
\newacro{CBM}{conduction band minimum}
\newacro{VDOS}{vibrational density of states}
\newacro{BZ}{Brillouin zone}
\newacro{DFT}{density-functional theory}
\newacro{VASP}{Vienna ab-initio simulation package}
\newacro{PAW}{projector-augmented wave}
\newacro{PBE}{Perdew-Burke-Ernzerhof}
\newacro{TS}{Tkatchenko-Scheffler}
\newacro{DFPT}{density functional perturbation theory}
\newacro{SOC}{spin-orbit coupling}
\newacro{TDS}{time-domain \unit{THz} spectroscopy }

\DeclareSIUnit{\wavenumber}{\unit{\cm^{-1}}}
\DeclareSIUnit{\rpm}{rpm}

\begin{document}
	\author{Sebasti\'{a}n Caicedo-D\'{a}vila}
	\affiliation{Physics Department, TUM School of Natural Sciences, Technical University of Munich, 85748 Garching, Germany}
	
	\author{Adi Cohen}
	\affiliation{Department of Chemical and Biological Physics, Weizmann Institute of Science, Rehovot 76100, Israel}
	
	\author{Silvia G. Motti}
	\affiliation{Clarendon Laboratory, Department of Physics, University of Oxford, Parks Road, Oxford, OX1 3PU, United Kingdom}
	\affiliation{School of Physics and Astronomy, Faculty of Engineering and Physical Sciences, University of Southampton, University Road, Southampton SO17 1BJ, United Kingdom}
	
	\author{Masahiko Isobe}
	\affiliation{Max Planck Institute for Solid State Research, 70569 Stuttgart, Germany}
	
	\author{Kyle M. McCall}
	\email{Present address: Department of Materials Science and Engineering, University of Texas at Dallas, 75080 Richardson, TX, USA}
	\affiliation{Laboratory of Inorganic Chemistry, Department of Chemistry and Applied Biosciences, ETH Zurich, CH-8093 Zürich, Switzerland.}
	\affiliation{Laboratory for Thin Films and Photovoltaics, EMPA - Swiss National Laboratories for Materials and Technology, CH-8600 Dübendorf, Switzerland}

 	\author{\textcolor{black}{Manuel Grumet}}
	\affiliation{Physics Department, TUM School of Natural Sciences, Technical University of Munich, 85748 Garching, Germany}
	
	\author{Maksym V. Kovalenko}
	\affiliation{Laboratory of Inorganic Chemistry, Department of Chemistry and Applied Biosciences, ETH Zurich, CH-8093 Zürich, Switzerland.}
	\affiliation{Laboratory for Thin Films and Photovoltaics, EMPA - Swiss National Laboratories for Materials and Technology, CH-8600 Dübendorf, Switzerland}
	
	\author{Omer Yaffe}
	\affiliation{Department of Chemical and Biological Physics, Weizmann Institute of Science, Rehovot 76100, Israel}
	\author{Laura M. Herz}
	\affiliation{Clarendon Laboratory, Department of Physics, University of Oxford, Parks Road, Oxford, OX1 3PU, United Kingdom}
	\affiliation{TUM Institute for Advanced Study, Technische Universität München, 85748 Garching, Germany}
	
	\author{Douglas H. Fabini}
	\email{dhfabini@gmail.com}
	\affiliation{Max Planck Institute for Solid State Research, 70569 Stuttgart, Germany}
	\affiliation{Department of Chemistry, Massachusetts Institute of Technology, 02139 Cambridge, MA, USA}
	
	\author{David A. Egger}
	\email{david.egger@tum.de}
	\affiliation{Physics Department, TUM School of Natural Sciences, Technical University of Munich, 85748 Garching, Germany}
	
	\title{Disentangling the Effects of Structure and Lone-Pair Electrons\\in the Lattice Dynamics of Halide Perovskites}
	
	
	\begin{abstract}
		\noindent Metal halide perovskites have shown great performance as solar energy materials, but their outstanding optoelectronic properties are paired with unusually strong anharmonic effects. It has been proposed that this intriguing combination of properties derives from the ``lone pair'' \textcolor{black}{6$s^2$} electron \textcolor{black}{configuration} of the \textcolor{black}{Pb$^{2+}$} cations, \textcolor{black}{and associated weak pseudo-Jahn--Teller effect,} but the precise impact of this chemical feature remains unclear.
		Here we show that in fact \textcolor{black}{an $ns^2$ electron configuration} is not a prerequisite for the strong anharmonicity and low-energy lattice dynamics encountered in this class of materials.
		We combine X-ray diffraction, infrared and Raman spectroscopies, and first-principles molecular dynamics calculations to directly contrast the lattice dynamics of \csb~with those of \cpb, two compounds which bear close structural similarity but with the former \textcolor{black}{lacking the propensity to form} lone pairs on the \textcolor{black}{5$s^0$} octahedral \textcolor{black}{cation}. 
		We exploit low-frequency diffusive Raman scattering, nominally symmetry-forbidden in the cubic phase, as a fingerprint to detect anharmonicity and reveal that low-frequency tilting occurs irrespective of \textcolor{black}{octahedral cation electron configuration}.
		This work highlights the key role of structure in perovskite lattice dynamics, providing important design rules for the emerging class of soft perovskite semiconductors for optoelectronic and light-harvesting devices.
	\end{abstract}
	
	\maketitle
	\noindent\acp{HaP} with formula \ce{AMX3} generated enormous research interest because of their outstanding performance in optoelectronic devices, most notably in efficient solar cells.~\cite{snaith2013,correa-baena2017,nayak2019}
	These compounds are highly unusual among the established semiconductors because they feature an intriguing combination of properties.
	Strong anharmonic fluctuations~\cite{beecher2016,whalley2016,carignano2017} in these soft materials appear together with optoelectronic characteristics that are favorable for technological applications.~\cite{brenner2016,Egger2018} 
	This confluence raised puzzling questions regarding the microscopic characteristics of the materials and the compositional tuning of their properties alike.
	On the one hand, the soft anharmonic nature of the \ac{HaP} structure may be beneficial in self-healing mechanisms of the material,\cite{ceratti2018,cahen2021,parida2023} allowing for low-energy synthesis routes in their fabrication.
	On the other hand, pairing of anharmonic fluctuations and optoelectronic processes for key quantities of HaPs, \textit{e.g.}, band gaps~\cite{patrick2015,wiktor2017,lanigan-atkins2021,seidl2023}, optical absorption profiles~\cite{gehrmann2019,gehrmann2022,wu2019}, and charge-carrier mobilities,~\cite{Egger2018,mayers2018,lacroix2020,iaru2021,schilcher2021,lai2022,zhang2022,schilcher2023} exposed incomplete microscopic rationales for the fundamental physical processes involved in solar-energy conversion.
	Established materials design rules are now being challenged by these observations, opening a gap in our protocols for making improved compounds.\\
	\noindent Significant efforts are now underway to discern the chemical effects giving rise to these remarkable properties of \acp{HaP}.
	Because lattice dynamical and optoelectronic properties appear both to be special and coupled in unusual ways, a common origin in chemical bonding could underlie these phenomena.
	In this context, an interesting chemical feature is that the octahedral cations in these compounds often bear an $ns^2$ electron configuration (\textit{e.g.}, \ce{Pb^2+} with configuration [Xe]6\textit{s}$^2$), which is not present in many other semiconductors.~\cite{fabini2020}
	This particular aspect of \acp{HaP} \textcolor{black}{leads to a ``strong'' or ``weak'' pseudo-Jahn--Teller (PJT) effect,}\cite{VanderVorst1980,Maaskant1991,Bersuker2013} \textcolor{black}{depending on the particulars of cation and anion composition and chemical pressure.}
    \textcolor{black}{The ``weak'' case influences local structure,~\cite{fabini2016,laurita2017} lattice dynamics~\cite{gao2021} and ionic dielectric responses,}~\cite{yaffe2017,fabini2016,fabini2020,huang2022} \textcolor{black}{while the ``strong'' case additionally results in the formation of a stereochemically-expressed electron lone pair and impacts average crystal structures.}\cite{walsh2011,smith2015,radha2018,gao2021} 
	\textcolor{black}{The weak PJT effect associated with 6$s^2$ Pb$^{2+}$ coordinated by heavy halides} plays a role in optoelectronic properties of these materials: its influence on the dielectric function can modify the Coulomb screening that is relevant for small exciton binding energies, reduced recombination rates and other key properties of \acp{HaP}.~\cite{du2014,herz2018} \\
	\noindent Confluences of the \textcolor{black}{propensity for lone-pair formation} with structural and lattice-dynamical properties were investigated in previous work exploring the chemical space of \acp{HaP}.
	Gao \textit{et al.}~\cite{gao2021} found an inverse relationship between the Goldschmidt tolerance factor, $t$,~\cite{goldschmidt1926} and anharmonic octahedral tilting motions.
	Similarly, Huang \textit{et al.} varied the A-site cation to explore interrelations of chemical, structural, and dynamical effects in \acp{HaP},~\cite{huang2022} reporting $t$-induced modulations of octahedral tiltings and \textcolor{black}{lone-pair} stereoactivity. 
	A recent study by several of the present authors found that \ce{Cs2AgBiBr6} lacks some expressions of lattice anharmonicity found in other \ac{HaP} variants.\cite{Cohen2022}
	Because every other octahedral cation (\ce{Ag^+}\textcolor{black}{, 4$d^{10}$}) cannot form a lone pair in this compound, \textcolor{black}{this raised the possibility that} changing the electron configuration of the cations may also suppress certain aspects of the lattice dynamics in \acp{HaP}.
	{Taken together, previous work assigned a central role of the \textcolor{black}{$ns^2$ electron configuration and associated PJT effect} in the anharmonic lattice dynamics of \acp{HaP} in addition to their established effect on the electronic structure and dielectric screening.}
	However, exploring the chemical space of \acp{HaP} in this way simultaneously changes their structures. 
	Therefore, isolating the convoluted occurrences of \textcolor{black}{cation lone-pair formation propensity} and purely structurally-determined changes in the lattice dynamics of \acp{HaP} remained challenging, making an assessment of the precise impact of \textcolor{black}{chemical bonding} on anharmonicity in these soft semiconductors largely inaccessible. \\
	\noindent Here, we address this issue and show that \textcolor{black}{an $ns^2$ cation compatible with lone-pair formation} is not required for the strong anharmonicity in the low-energy lattice dynamics of soft \ac{HaP} semiconductors.
	We disentangle structural and chemical effects in the lattice dynamics of \acp{HaP} by comparing the well-known \cpb~with the far less studied \csb. 
	Both exhibit almost identical geometrical and structural parameters, but \csb~\textcolor{black}{exhibits a negligible PJT effect on the octahedral Sr$^{2+}$ site, owing to weaker vibronic coupling to degenerate excited states of appropriate symmetry which lie higher in energy than in the Pb$^{2+}$ case}, allowing separation of the effects of the \textcolor{black}{$ns^2$ electron configuration} and the geometry on the lattice dynamics in a direct manner.
	Combining electronic structure and \ac{MD} calculations with \ac{XRD}, \ac{IR} and Raman spectroscopies, we assess a key fingerprint of vibrational anharmonicity, \textit{i.e.}, the Raman central peak, which is a broad peak towards zero frequency in the Raman spectrum resulting from diffusive inelastic scattering \cite{yaffe2017,ferreira2020,fabini2020,sharma2020,gao2021,huang2022,menahem2023}.
	While the electronic structure and dielectric properties of \cpb~and \csb~are very different, their vibrational anharmonicities are found to be remarkably similar.
	In particular, the crucial dynamic octahedral tiltings giving rise to the Raman central peak are still present even in the absence of \textcolor{black}{$ns^2$ octahedral cations} in \csb.
	Our results provide microscopic understanding of precisely how the \textcolor{black}{propensity for lone-pair formation} influences the anharmonic octahedral tiltings that dynamically break the average cubic symmetry in both compounds, and rule out the \textcolor{black}{weak PJT associated with the $ns^2$ main-group cations} as the sole reason for the appearance of such anharmonicity in soft \acp{HaP}.
	These findings are important for chemical tuning of \acp{HaP} needed for new materials design.
	\section*{Results}\label{sec:Results}
\begin{figure}
		\centering
		\includegraphics[width=\linewidth]{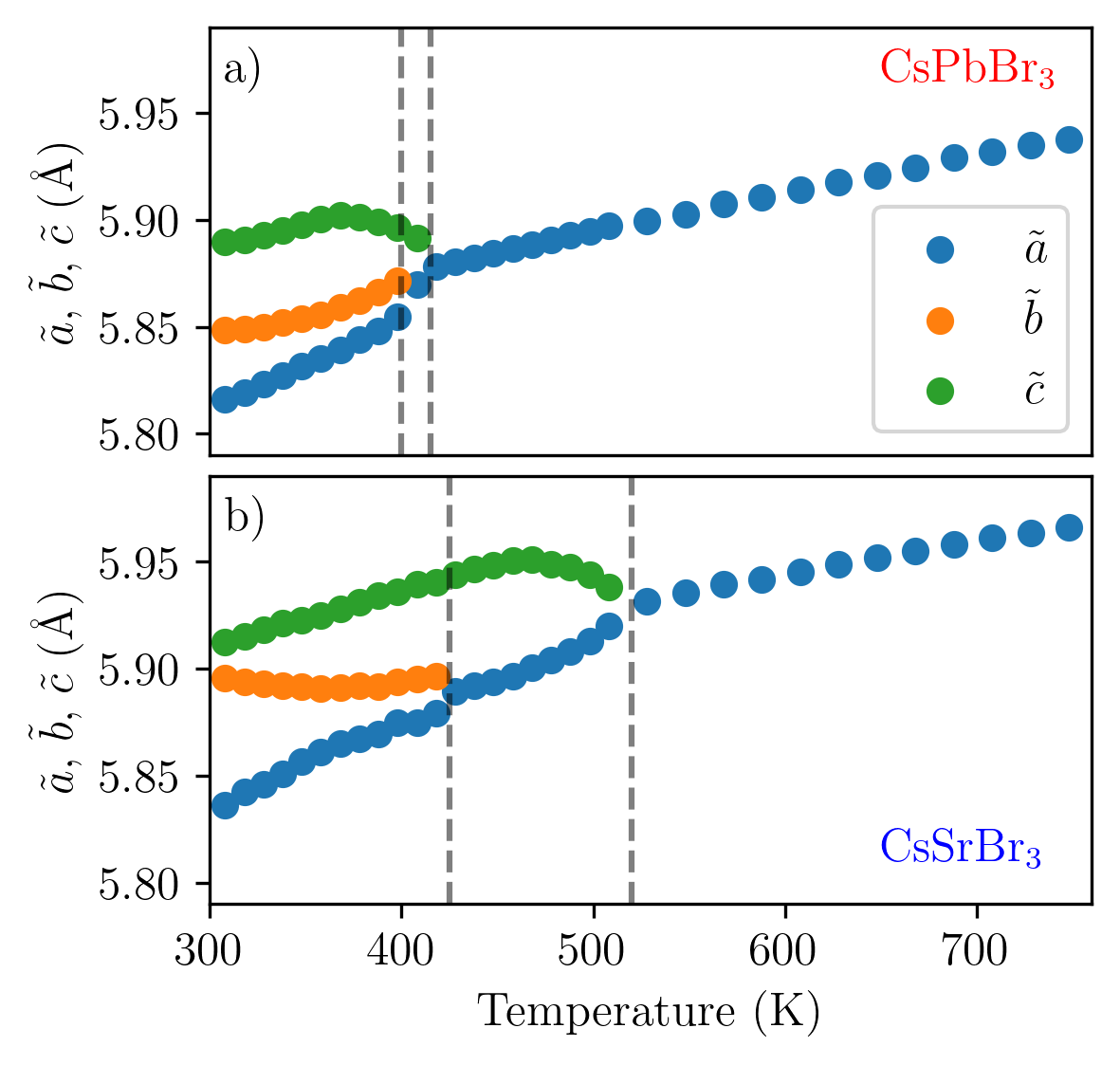}
		\caption{
			\textbf{Structural properties.} Temperature-dependent lattice parameters of \cpb~(panel a) and \csb~(panel b) determined by \ac{XRD} throughout the orthorhombic---tetragonal---cubic phases. 
			We show reduced lattice parameters $\tilde{a}$, $\tilde{b}$ and $\tilde{c}$ for better visualization, \textcolor{black}{with the orthorhombic phase expressed in the $Pbnm$ setting}.
			Dashed vertical lines indicate phase-transition temperatures.
			Error bars from Pawley fitting are smaller than the markers and are omitted.
		}
		\label{fig:XRD(T)}
	\end{figure}

\begin{figure}[b]
		\centering
		\includegraphics[width=\linewidth]{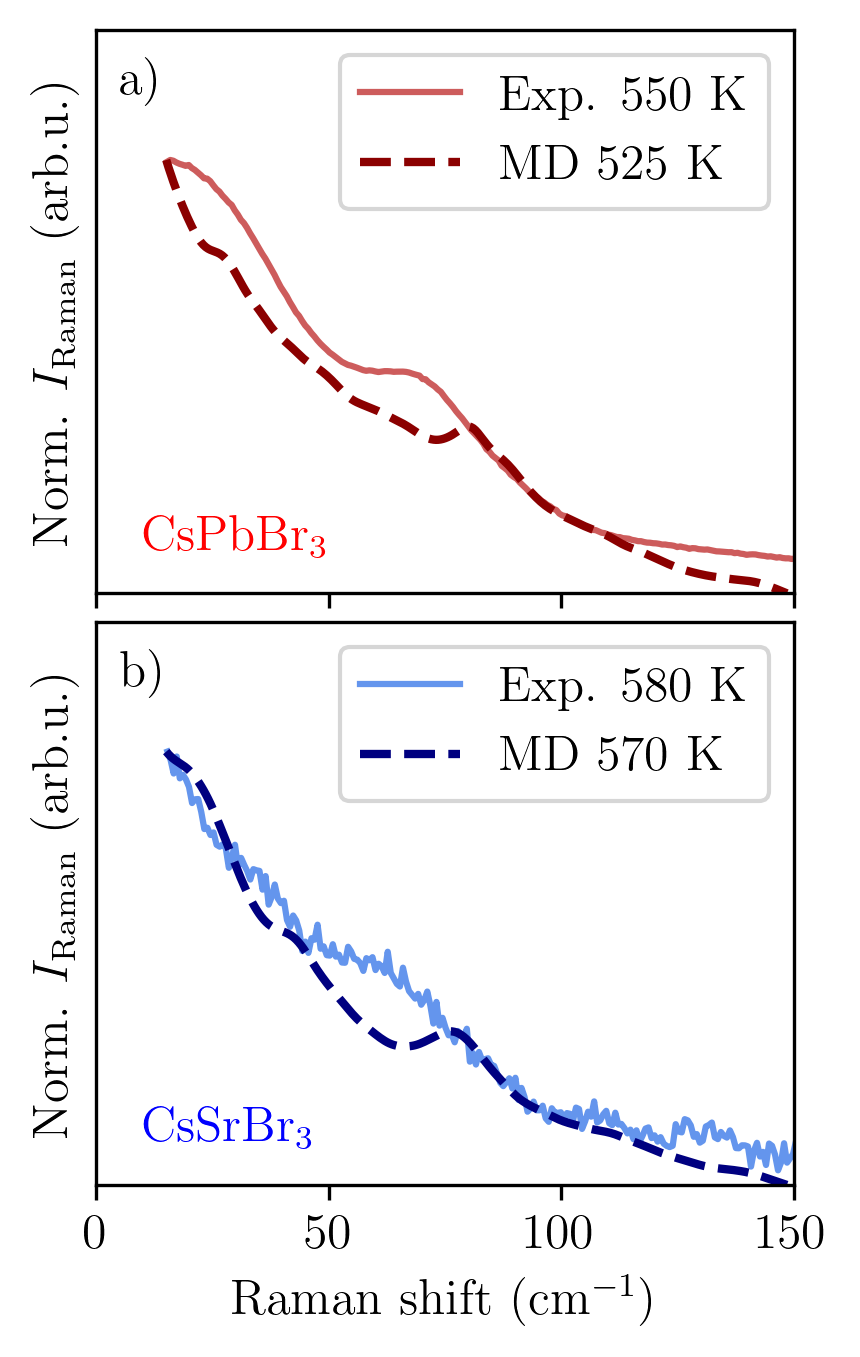}
		\caption{\textbf{Lattice dynamics at higher temperature.}
			Raman spectra of \cpb~\textcolor{black}{(panel a)} and \csb~\textcolor{black}{(panel b)} in the high-temperature cubic phase \textcolor{black}{measured experimentally and calculated using DFT-MD.} 
			The central peak appears for both compounds \textcolor{black}{in the experiments and computations despite significant differences in bonding: [PbBr$_6$]$^{4-}$ is proximate  to lone-pair formation (\textit{i.e.}, exhibits a ``weak'' PJT effect),\cite{Bersuker2013} while PJT effects associated with [SrBr$_6$]$^{4-}$} are negligible.
		}
		\label{fig:HT-Raman}
	\end{figure}

\begin{figure}
		\centering
		\includegraphics[width=\linewidth]{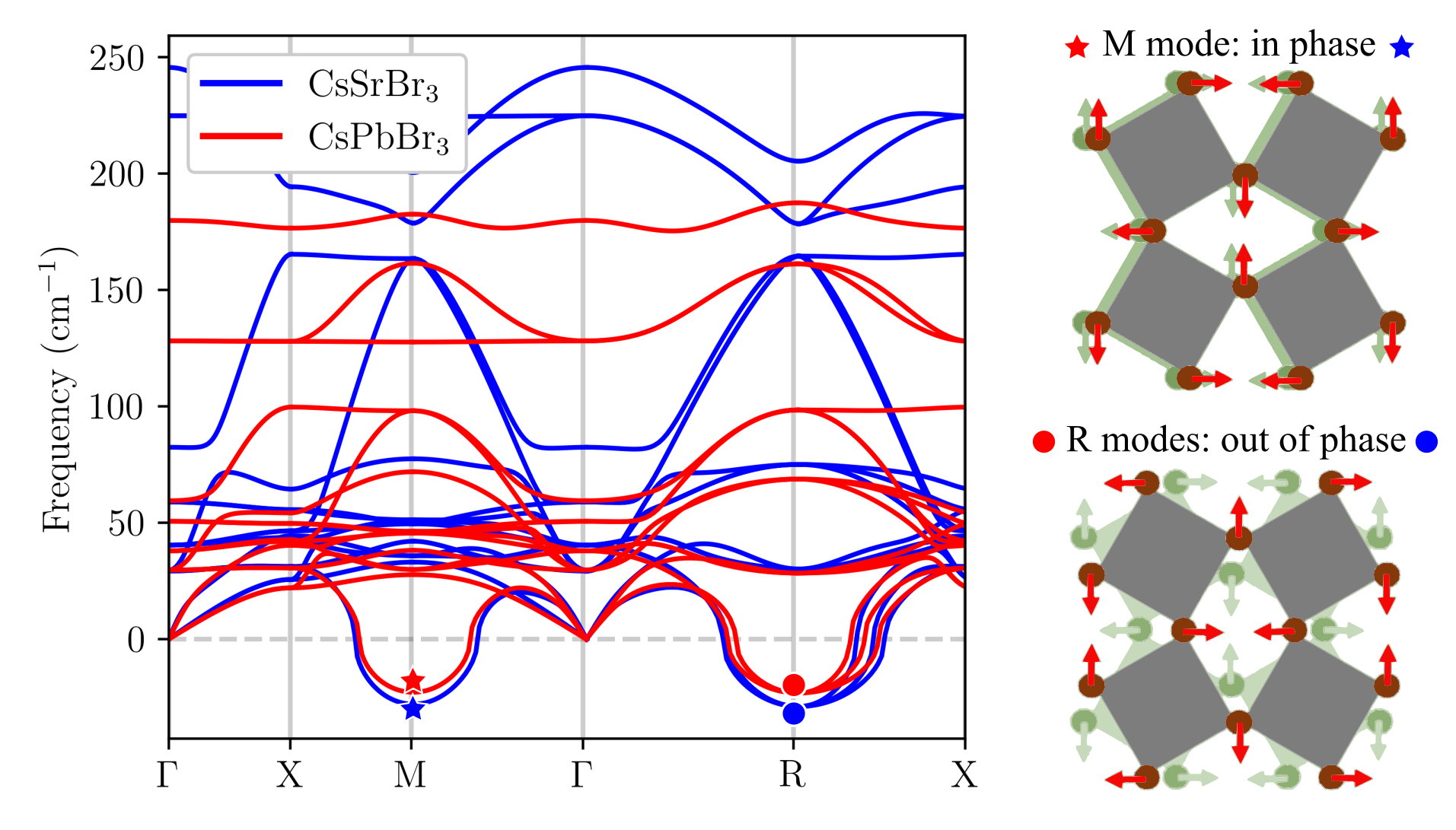}
		\caption{\textbf{Dynamic instabilities in the lattice dynamics.}
			Harmonic phonon dispersion of cubic \cpb~and \csb~showing the dynamic instabilities in the high-temperature, cubic phase of both compounds. 
			The imaginary modes at the M and R points are the in-phase and out-of-phase tilting depicted on the right panels. 
			The tilting modes are almost identical for \csb~and \cpb.
		}
		\label{fig:phonon_dispersion}
	\end{figure}

	\begin{center}
		\textbf{Electronic structure and bonding} \\
	\end{center}
	\begin{figure}[b]
		\centering
		\includegraphics[width=\linewidth]{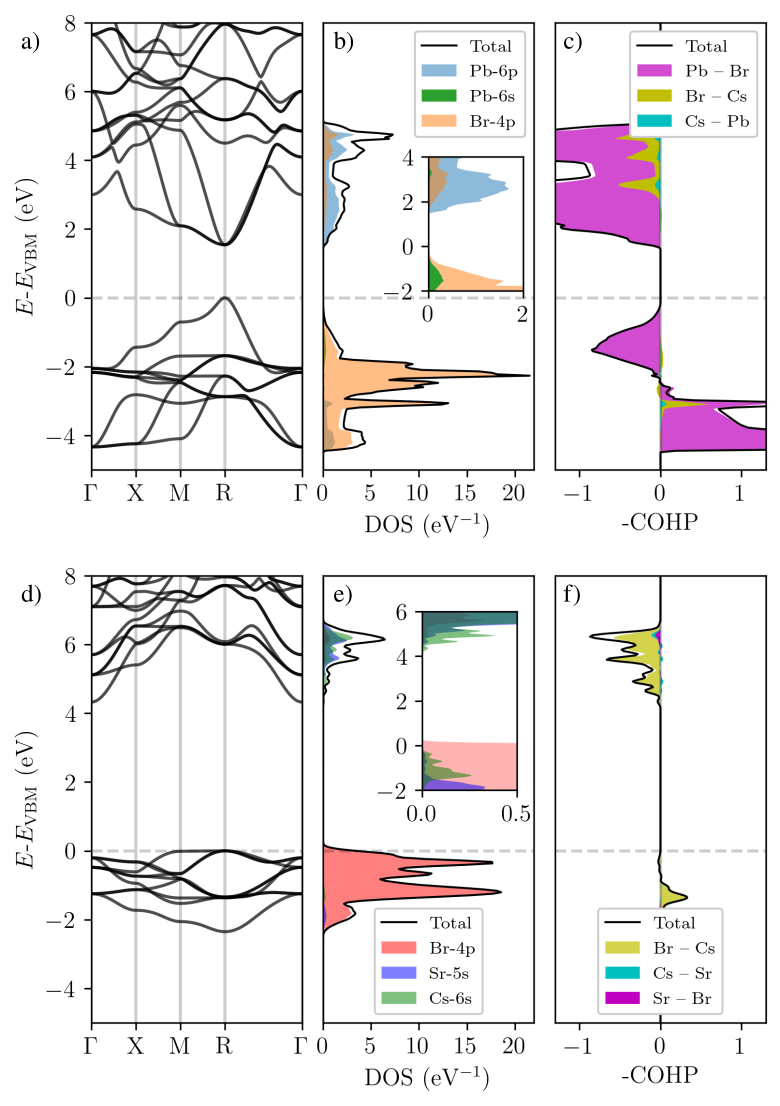}
		\caption{\textbf{Electronic structure.} DFT-computed electronic band structure of cubic \cpb~(panel a) and corresponding total and projected density of states (DOS, panel b) and crystal-orbital Hamilton population (COHP, panel c). 
			Panels d---f show the same data for \csb.
		}
		\label{fig:band_structure}
	\end{figure}%
	\noindent We first investigate the electronic structure and bonding of \cpb~and \csb~using \ac{DFT}.
	Figure~\ref{fig:band_structure} shows their band structure, total and projected \ac{DOS}, as well as the total and projected \ac{COHP} of the high-temperature cubic phases of \cpb~and \csb.
	The electronic band structure and bonding of \cpb~were extensively investigated before:~\cite{goesten2018} 
	the \ac{CBM} is formed by anti-bonding interactions (positive \ac{COHP} in Figure~\ref{fig:band_structure}c) between Pb-6\textit{p} and Br-4\textit{p}/Br-4\textit{s} orbitals, while the \ac{VBM} is formed by anti-bonding interactions between Br-4\textit{p} and Pb-6\textit{s} orbitals.\\
	\noindent The electronic structure of \csb~exhibits entirely different characteristics,~\cite{fabini2020,straus2022} especially a much larger band gap and weaker covalent interactions.
	Notably, the magnitude of the COHP is significantly reduced with respect to that of \cpb, indicating much greater ionicity, and the \ac{COHP} is almost entirely recovered by interactions between Cs and Br. Importantly, all bands derived from antibonding interactions between Sr-5\textit{s} and Br-4\textit{p}/Br-4\textit{s} are empty due to the electron configuration of \ce{Sr^{2+}} ([Kr]), and there is no potential for lone pair formation on \ce{Sr^{2+}}.\\
	\noindent A manifestation of the lack of \textcolor{black}{$ns^2$ cations} in \csb~is that there is no cross-gap hybridization of the halide valence orbitals.
	By contrast, Br-4\textit{p} orbitals hybridize with Pb-6\textit{p} across the gap of \cpb~(see the p\ac{COHP} in Figure~\ref{fig:band_structure}c).
	This leads to large Born effective charges, \textit{i.e.}, large changes in the macroscopic polarization upon ionic displacements~\cite{du2010,sun2017,ran2018,kang2018} reported in Table \ref{tab:BORN-dielectric}, which for \cpb~are more than double the formal charge of Pb (+2) and Br (-1) and much larger than the corresponding values for \csb.
	Similarly, there is also a larger electronic contribution to the dielectric response in \cpb~and it features a larger value of the dielectric function at the high-frequency limit ($\varepsilon_\infty$) compared to \csb.\\
	\begin{table}
		\centering
		\caption{\textbf{Dielectric properties of cubic \ce{CsMBr3}.} Dielectric constant in the high-frequency limit with respect to the optical phonon mode frequencies, $\varepsilon_\infty$, and Born effective charges, $Z_{i}^{*}$, of cubic \cpb~and \csb~as calculated by \ac{DFT}. 
			We report $Z_{\mathrm{Br}}^{*}$ for the Br bonded with Pb/Sr along the $z$ axis.}
		\begin{tabular}{c|c|c|c|c}
			\multirow{2}*{Compound} & \multirow{2}*{$\varepsilon_\infty$}	& \multirow{2}*{$Z_{\mathrm{Cs}}^{*}$} & \multirow{2}*{$Z_{\mathrm{M-site}}^{*}$} & $Z_{\mathrm{Br}}^{*}$ \\
			&								&			&		&	$(xx,yy,zz)$\\
			\hline\hline
			CsPbBr$_3$				&			5.39				&	1.38	&	4.33	&	 (-0.63, -0.63, -4.46)\\
			CsSrBr$_3$				&			3.02				&	1.35	&	2.43	&	 (-0.91, -0.91, -1.97)
		\end{tabular}
		\label{tab:BORN-dielectric}
	\end{table}
	\begin{center}
		\textbf{Structural properties and phase transitions}\label{subsec:phase-transitions}\\
	\end{center}
	\noindent In spite of the markedly different electronic structure and bonding characteristics, \csb~and \cpb~exhibit the same high-temperature cubic crystal structure ($Pm\bar{3}m$) and very similar lattice parameters (see Supplementary Information).
	One can rationalize this through the nearly identical ionic radii of \ce{Pb^2+} and \ce{Sr^2+} (\qtylist[list-units=single]{119;118}{\pico\meter}) and the resulting Goldschmidt factors for the compounds (0.862 and 0.865).
	Furthermore, both materials exhibit the same sequence of structural phase transitions from the high-temperature cubic to the low-temperature orthorhombic phase (with an intermediate tetragonal phase), as shown by temperature-dependent lattice parameters in Figure \ref{fig:XRD(T)} that were determined \textit{via} \ac{XRD}.
	The cubic-to-tetragonal phase transition temperature of \csb~($\sim$\SI{520}{\kelvin}) is noticeably higher than that of \cpb~($\sim$\SI{400}{\kelvin})~\cite{rodova2003,stoumpos2013} and slightly higher ($\sim$\SI{10}{\kelvin}) than that reported for Eu-doped \csbe~5\%.~\cite{loyd2018}
	The volumetric thermal expansion coefficient ($\alpha_{V}$) of \csb~($\sim$\SI{1.32E-4}{\kelvin^{-1}} at \SI{300}{\kelvin}) is large and similar to that of \cpb~($\sim$\SI{1.29E-4}{\kelvin^{-1}}, see the Supplementary Information for details), in good agreement with the one reported for \csbe.~\cite{loyd2018}
	Just as for other inorganic \acp{HaP}, $\alpha_V$ of \csb~slightly decreases with temperature.~\cite{fabini2016a,schueller2018}
	The similarity of geometric factors and structural phase transitions suggests that the octahedral tilting dynamics in \csb~might be similar to those in \cpb, which contrasts with their markedly different electronic structure, and prompts a deeper investigation of the impact of the \textcolor{black}{$ns^2$ cations} on structural dynamics. \\
	\begin{figure*}
		\centering
		\includegraphics[width=0.7\textwidth]{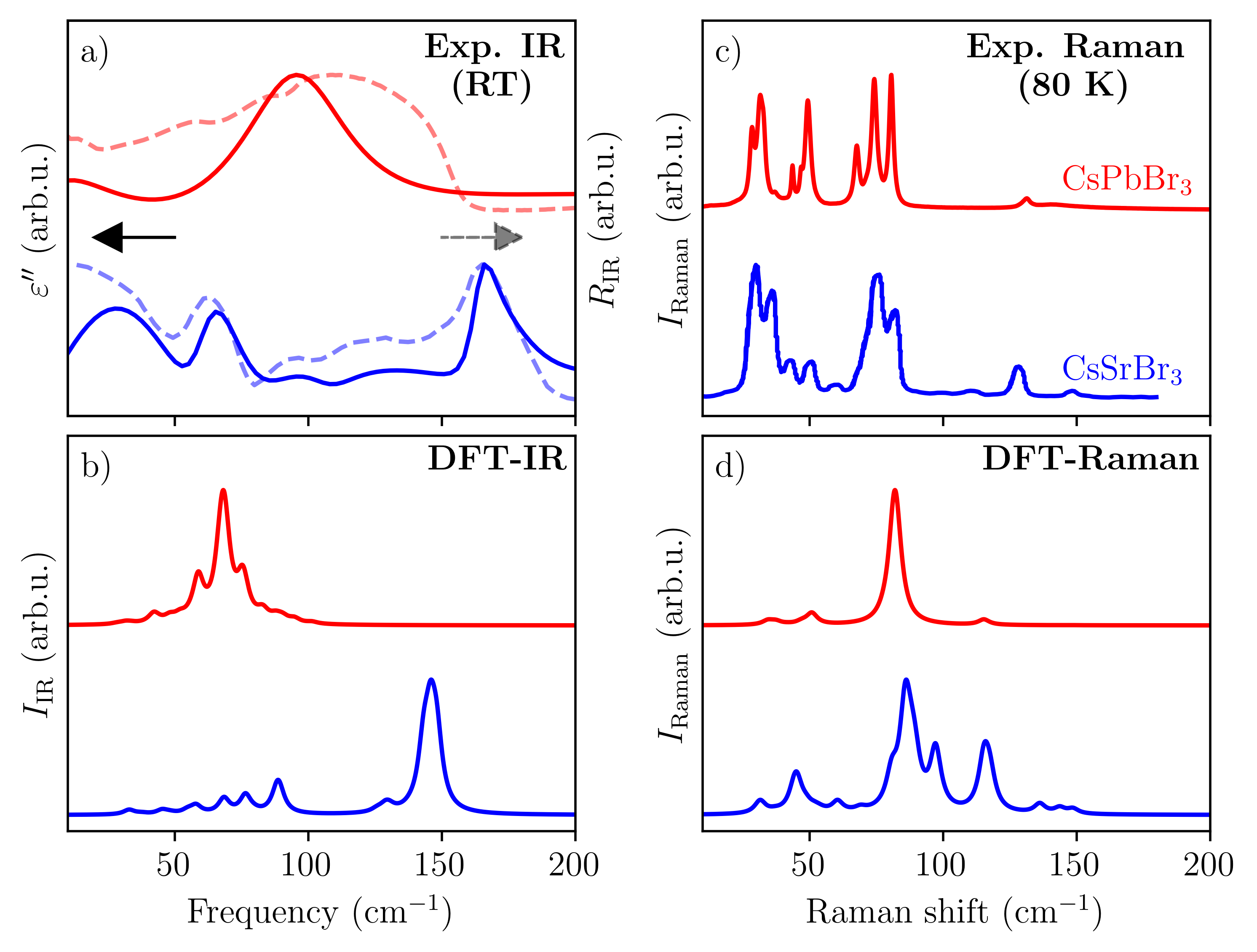}
		\caption{\textbf{Lattice dynamics at lower temperatures.}
			a) \ac{IR}-reflectivity spectra (dashed curves) and fitted imaginary part of the dielectric function (solid curves, see Supplementary Information for details) of \cpb~and \csb~measured at room temperature.
			b) \ac{DFT}-calculated \ac{IR}-absorption spectra within the harmonic approximation for the orthorhombic phases.
			c) Raman spectra of orthorhombic \cpb~and \csb~measured at \SI{80}{\kelvin}. 
			d) \ac{DFT}-calculated Raman spectra of both compounds within the harmonic approximation for the orthorhombic phases.
		}
		\label{fig:LT_spectra}
	\end{figure*}
	\begin{center}
		\textbf{Lower-temperature lattice dynamics}\label{subsec:LT-lattDyn}\\
	\end{center}
	\noindent We conduct \ac{IR} and Raman spectroscopy at different temperatures as well as \ac{DFT}-based harmonic-phonon calculations. 
	The measured \ac{IR} spectra show that the dominant \csb~features are blue-shifted compared to those of \cpb~(see Figure~\ref{fig:LT_spectra}a). 
	Indeed, our \ac{DFT} calculations of \ac{IR} activities find a significant softening of the infrared-active TO modes in \cpb~compared to those in \csb~(see Figure~\ref{fig:LT_spectra}b): the most prominent \ac{IR}-active TO mode in \cpb~and \csb~appears at $\sim$\SIlist{68;146}{\wavenumber}, respectively, corresponding to the same irreducible representation ($B3u$) with similar eigenvectors (see Supplementary Information) in each system.
	This is in line with the \textcolor{black}{theory of weak PJT effects in general}\cite{Bersuker2013} \textcolor{black}{and expectations for lone pairs in particular, with significant softening of \textit{ungerade} modes in \cpb~that would correspond to lone-pair formation in the strong PJT case relative to those in \csb. Notably, this softening is primarily driven by differences in bonding rather than} the difference in the atomic masses (see Supplementary Information).\\
	Moreover, the LO/TO splitting is enhanced in \cpb~compared to in \csb~and the LO phonon modes are hardened.
	Related to this, the \cpb~\ac{IR} spectrum exhibits a broad feature which is known as the Reststrahlen band as has been reported before for MA-based \acp{HaP}.\cite{sendner2016}
	This particular effect results in near-zero transmission through the material in a frequency range between the TO and LO modes, represented by high \ac{IR} intensity values, and occurs in polar materials with larger Born-effective charges. 
	Because the TO modes are softened and the LO modes hardened in \cpb~compared to \csb, and because the latter is less polar (\textit{cf.} Table~\ref{tab:BORN-dielectric}), the absence of the \textcolor{black}{$ns^2$ cations} leads to a much less pronounced, blue-shifted Reststrahlen band appearing in a smaller frequency window in \csb~(see Figure~\ref{fig:LT_spectra}a, and Supplementary Information).\\
	Figure \ref{fig:LT_spectra}c shows the \SI{80}{\kelvin} Raman spectra of \cpb~and \csb, which are in good agreement with the Raman activities calculated for harmonic phonons (Figure~\ref{fig:LT_spectra}d).
    \textcolor{black}{Specifically, the experimental spectrum of CsPbBr$_3$ in Figure~\ref{fig:LT_spectra}b finds three broader features at frequencies below and one weaker-intensity feature at frequencies above 100\,cm$^{-1}$. 
    Conversely, CsSrBr$_3$ exhibits a structured feature around 50\,cm$^{-1}$, a pronounced signal close to 100\,cm$^{-1}$, and then a series of weaker intensities between 100--150\,cm$^{-1}$.
    While the DFT-computed Raman activities calculated in the harmonic approximation are in broad agreement with these findings (see Figure~\ref{fig:LT_spectra}d), we note a slightly larger deviation of approximately 20\,cm$^{-1}$ for the higher-frequency peak in CsPbBr$_3$.
    These findings lead us to conclude that unlike in} \ac{IR}, the Raman spectrum of \csb~exhibits no substantial energy shifts with respect to \cpb. 
	Computing the phonon \ac{DOS} for the orthorhombic phase of both compounds with \ac{DFT} (see Supplementary Information), we find that they exhibit similar phonon \ac{DOS} below \SI{100}{\wavenumber}, \textit{i.e.}, in the region of most of the Raman-active modes. The similar phonon \ac{DOS} and the contributions of the M-site at low frequencies explain the limited shift of the \csb~Raman spectrum\textcolor{black}{, which might be surprising at first sight given the different atomic masses of Sr and Pb.}
	Above this range, \cpb~exhibits few vibrational states while \csb~shows its most pronounced phonon DOS peaks, which correspond well with the strongest \ac{IR} mode calculated from the harmonic approximation.\\
	\begin{center}
		\textbf{High-temperature lattice dynamics}\label{subsec:HT-lattDyn}\\
	\end{center}
	\noindent A key signature of vibrational anharmonicity in \acp{HaP} at higher temperatures is the Raman central peak.~\cite{yaffe2017,ferreira2020,fabini2020,sharma2020,gao2021,huang2022,menahem2023}
	We use this feature that is nominally symmetry-forbidden in the cubic phase as a fingerprint to directly investigate how the \textcolor{black}{propensity for cation lone-pair formation or lack thereof} determines anharmonicity in these materials, using Raman spectroscopy and \ac{DFT}-based \ac{MD} simulations.
	Interestingly, a central peak also appears in the high-temperature Raman spectrum of \csb~(see Figure~\ref{fig:HT-Raman} and  Supplementary Information for full temperature range)\textcolor{black}{. 
    We note that differences in Raman intensity imply that the scattering cross-section of CsSrBr$_3$ is notably weaker than that of CsPbBr$_3$, which is due to its significantly higher bandgap and weaker dielectric response at the Raman excitation wavelength (785~nm) and because a powder sample of \csb~has been used for which scattering of light in the back-scattering direction is considerably lower.
	The presence of a central peak} in \csb~shows that \textcolor{black}{local fluctuations associated with a cation lone-pair} are not required for the low-frequency diffusive Raman scattering and anharmonicity to occur.
	This result, together with the identical phase-transition sequences of both materials (see Figure~\ref{fig:XRD(T)}), led us to investigate the role of tilting instabilities in \csb~and \cpb.\\
	\noindent We first calculate \textcolor{black}{the Raman spectrum for both compounds using MD calculations (see Figure~\ref{fig:HT-Raman} and Methods section). 
    Remarkably, a central peak appears also in the MD-computed high-temperature Raman spectrum of \cpb~and \csb. 
    We find good agreement between experiment and theory, both showing a feature between 50--100\,cm$^{-1}$ in the Raman spectra of the two materials in addition to the central peak.\\
 Next, we compute }harmonic phonon dispersions of both compounds (see Figure~\ref{fig:phonon_dispersion}) and find these to be remarkably similar for cubic \csb~and \cpb~in the low frequency region, in line with the aforementioned similarities in the phonon DOS of the orthorhombic phase.
	Specifically, both compounds exhibit the same dynamic tilting instabilities at the edge of the \ac{BZ}, governed by in-phase (M point) and three degenerate out-of-phase (R point) rotations.
	These rotation modes are not only active in the phase transitions, but they also have been discussed to drive the dynamic disorder of halide perovskites.~\cite{armstrong1989,woodward1997,beecher2016,yang2017a,yang2020,lanigan-atkins2021} \\
	\begin{figure*}[t!]
		\centering
		\includegraphics[width=0.8\linewidth]{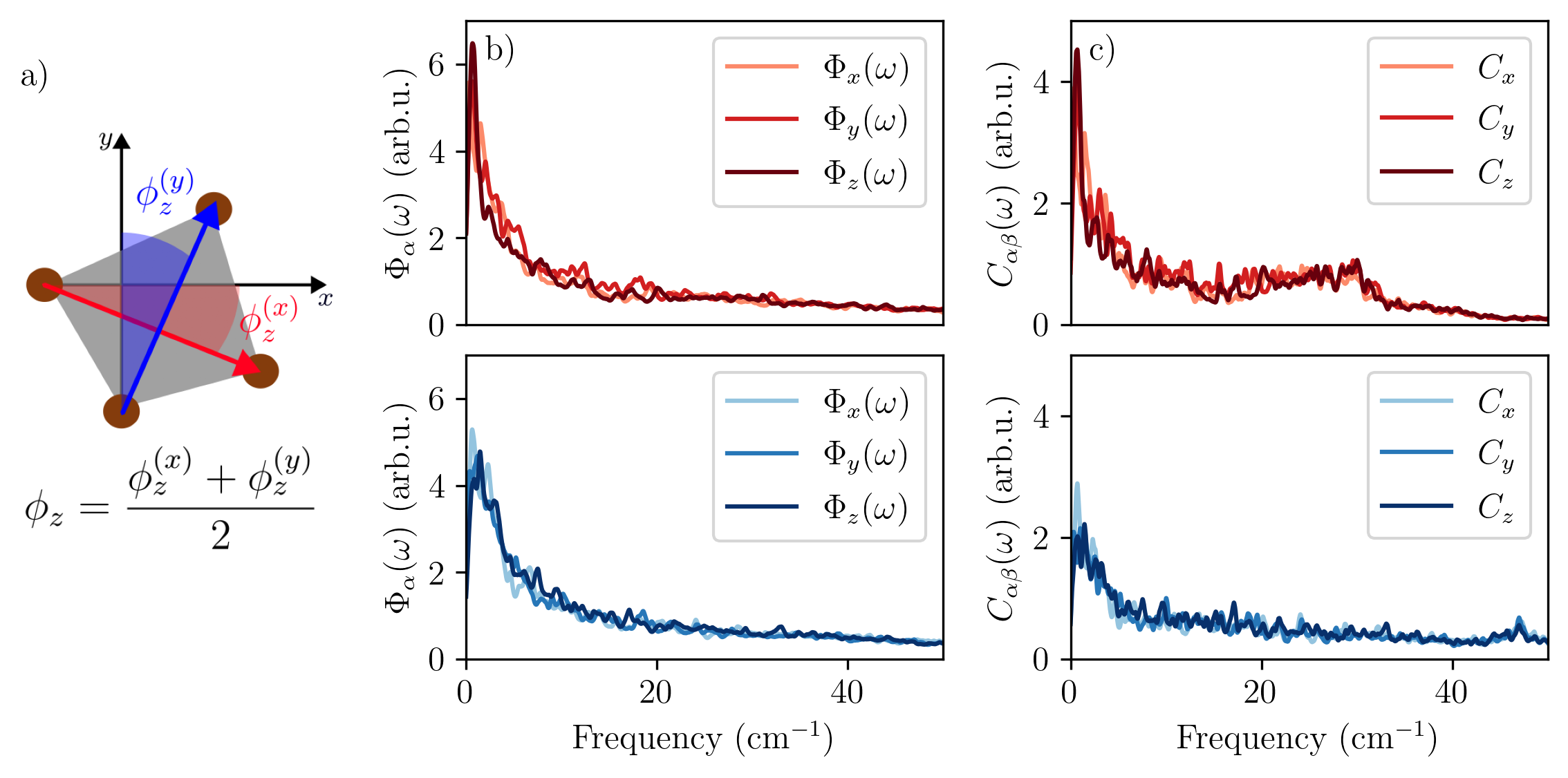}
		\caption{\textbf{Impact of \textcolor{black}{cation electron configuration} on octahedral dynamics at higher temperature.}
			a) Schematic representation of the \ce{MBr6} octahedron aligned along the \textit{z} Cartesian axis. The octahedral rotation angle around $z$, $\phi_z$, is defined as the average of the angles formed by the $x/y$ Cartesian axis and the vector connecting two in-plane Br atoms at opposing edges of the octahedron ($\phi_z^{(x)}$ in red and $\phi_z^{(y)}$ in blue). Note that a clockwise rotation is defined as positive and counter-clockwise as negative.
			b) Fourier transform of the octahedral rotation angle, $\Phi_{\alpha}(\omega)$, and c) cross-correlation between rotation angle and M-site displacement, $C_{\alpha\beta}(\omega)$, calculated using \ac{DFT}-\ac{MD} trajectories of cubic  \cpb~(upper panels, \SI{525}{\kelvin}) and \csb~(lower panels, \SI{570}{\kelvin}).
		}
		\label{fig:VDOS}
	\end{figure*}%
	\noindent \textcolor{black}{Finally, using the \ac{MD} trajectories of \cpb~and \csb~in the cubic phase,} we calculate the frequency-resolved dynamic changes of octahedral rotation angles, $\Phi_{\alpha}(\omega)$ (see Figure \ref{fig:VDOS} and Equation \ref{eq:ang-frequency} in the Methods Section).
	Figure \ref{fig:VDOS}b shows $\Phi_{\alpha}(\omega)$ for \cpb~and \csb~and indicates strong low-frequency tilting components in both \cpb~and \csb.
	Recently, a phenomenological model for the description of the temperature-dependent Raman spectra of cubic \acp{HaP} proposed the inclusion of a low-frequency anharmonic feature, which was associated with transitions between minima of a double-well potential energy surfaces~\cite{menahem2023} that correspond to different octahedral tiltings.~\cite{yang2017a,klarbring2019,bechtel2019,zhu2022}
	Our results confirm that substantial octahedral dynamics correspond to low-frequency features dynamically breaking the cubic symmetry in \cpb~and \csb.~\cite{beecher2016,sharma2020,lanigan-atkins2021,wiktor2023,baldwin2023}
	Interestingly, this low-frequency component appears irrespective of the presence of \textcolor{black}{$ns^2$ cations} and induces the formation of relatively long-lived (tens of ps) structural distortions (see Supplementary Information), which strongly deviate from the average cubic symmetry.
	This suggests that the dynamic deviations from the long-range, crystallographic structure enable the low-frequency Raman response without violating the selection rules.\\
	\noindent We investigate the impact of the \textcolor{black}{M-site chemistry} on octahedral tilting tendencies~\cite{gao2021} by computing the Fourier-transform of cross-correlations between rotation angles and M-site displacements, $C_{\alpha\beta}(\omega)$ (see Equation \ref{eq:Corr} in the Methods section).
	Larger values of $C_{\alpha\beta}$ generally indicate stronger coupling between octahedral rotations and Pb displacements.
	Absence of the \textcolor{black}{propensity for lone-pair formation} becomes evident in the low intensity of $C_{\alpha\beta}(\omega)$ for \csb~(Figure \ref{fig:VDOS}c), which is less than half of that of \cpb, especially at low-frequencies relevant for the slow, anharmonic, symmetry-breaking rotational features.
	This suggests that the presence of the \textcolor{black}{$ns^2$ cations} in \cpb~enhances the low-frequency octahedral tilting, in line with the literature.~\cite{gao2021}
    \textcolor{black}{
    M-site displacements and octahedral rotations are correlated because the latter is accompanied by changes of the Br-Pb-Br resonant network\cite{gehrmann2022} affecting the charge density in the vicinity of the M-site.
    While this effect is very weak in \csb~(see Supplementary Information),
	the non-zero $C_{\alpha\beta}$ for this case shows that the presence of \textcolor{black}{$ns^2$ cations} is not necessary to couple octahedral rotations and M-site displacements because the ions are still interacting through other types of interactions, \textit{e.g.}, electrostatically or due to Pauli repulsion.
    In CsPbF$_3$, the interaction of tilting and M-site displacements is strong enough to drive the adoption of an unusual tilt pattern.\cite{smith2015}
    }
    We speculate that the \textcolor{black}{lone-pair}-enhanced tilting could contribute to the fact that \cpb~has a lower tetragonal-to-cubic phase transition temperature compared to \textcolor{black}{that of} \csb.\\
	\begin{center}
		\textbf{Discussion}\label{subsec:Optoelectronics}\\
	\end{center}

	\noindent \textcolor{black}{We directly disentangled structural and chemical effects in \acp{HaP} by comparing \cpb~and \csb, two compounds with similar ionic radii and structural properties but entirely different orbital interactions that imbue \cpb~with the weak PJT effect common to technologically-relevant Pb perovskites and \csb~with negligible PJT effects.
    While the $ns^2$ configuration of the octahedral cations is paramount for the optoelectronic and dielectric properties of these materials, using the Raman central peak at higher temperatures as a fingerprint to detect anharmonicity we found it to appear also for \csb~with 5$s^0$ cations and to correlate with slow, anharmonic rotations of the octahedra.}
	Altogether, these findings demonstrate that the perovskite structure allows for anharmonic vibrational dynamics to occur, irrespective of the presence of \textcolor{black}{$ns^2$ cations with the propensity to form lone pairs}, which establishes this somewhat unusual behavior as a generic effect in this material class. 
    \textcolor{black}{We note that recent work by some of the present authors has investigated the commonalities and differences between oxide perovskites and \acp{HaP} in this context.}~\cite{menahem2023} \\
	\noindent Since octahedral dynamics impact the optoelectronic characteristics of these systems, our results have implications for synthesis of new \acp{HaP} with improved properties for technological applications. 
	For instance, Pb-Sr alloying has been proposed as a method to tune the band gap of \acp{HaP} for light emission and absorption applications.\cite{straus2022} Our work implies that such Sr alloying for tuning electronic and dielectric properties preserves the strongly anharmonic lattice dynamics.
    \textcolor{black}{{Furthermore, investigating related compounds with distinct electronic configurations on the octahedral cation, such as \ce{CsEuBr3}, may provide further insight about chemical trends in tuning of the \ac{HaP} properties.}}\\
    \noindent The relevance of these findings for material design strategies of \ac{HaP} compounds is \textcolor{black}{additionally affirmed} when putting our results in the context of previous work discussing anharmonic effects in this class of materials.
	Specifically, cubic \ce{CsPbBr3}, \ce{CsSnBr3}, \ce{CsGeBr3}, \ce{(CH3NH3)_{0.13}(CH3CH2NH3)_{0.87}PbBr3}, \ce{CH(NH2)2PbBr3}, and, here, \ce{CsSrBr3} are all reported to exhibit dynamic hopping between low symmetry minima on the potential energy surface.~\cite{yaffe2017,gao2021,huang2022,Reuveni2023} 
	By contrast, the high symmetry phase of \ce{Cs2AgBiBr6} is anharmonically stabilized and exhibits well-defined normal modes and a soft-mode transition on cooling.~\cite{Cohen2022}
	\ce{Cs2SnBr6}, on the other hand, lacks any phase transitions and similarly exhibits well-defined normal modes.\cite{Kaltzoglou2016}
	Where previously the strength of the \textcolor{black}{PJT effect associated with $ns^2$ cations} or the density of \textcolor{black}{such} cations appeared to be a plausible predictor of broad, nominally symmetry-forbidden Raman scattering resulting in a central peak, our work suggests that instead the differing symmetry in both the structure and the chemical bonding of metal halide perovskites and double-perovskites may be a controlling factor.
    \textcolor{black}{Notably, CsGeBr$_3$, which exhibits no octahedral tilting transitions~\cite{Thiele1987} and a broad Raman central peak in the cubic phase with a mode reflecting persistent pyramidal [GeBr$_3$]$^-$ anions,~\cite{gao2021} corresponds to the ``strong'' PJT~\cite{Bersuker2013} case:
    Stereochemically expressed cation lone pairs are evident in the low temperature average structure~\cite{Thiele1987} and in the local fluctuations of the cubic phase.\cite{gao2021}
    Dynamic symmetry-breaking giving rise to a broad Raman central peak is thus observed for three distinct bonding regimes with regard to pseudo-Jahn--Teller effects: strong PJT (CsGeBr$_3$),\cite{gao2021} weak PJT (CsPbBr$_3$ and others),\cite{yaffe2017} and negligible PJT (CsSrBr$_3$).}\\
	\noindent In conclusion, the n$s^2$ electron configuration in \acp{HaP} that can result in \textcolor{black}{formation of lone-pairs} is crucial to several favorable electronic features\cite{du2014,goesten2018,fabini2020} and gives rise to the elevated ionic dielectric response \textit{via} enhancement of Born effective charges.~\cite{du2010,du2014} 
	However, we found that presence of a \textcolor{black}{strong or weak PJT effect associated with $ns^2$ cations} is not necessary to produce dynamic symmetry-breaking of the sort that gives rise to broad, intense Raman scattering in the high temperature phases of \acp{HaP} and that has been associated with the unique optoelectronic properties in these compounds such as long charge-carrier lifetimes and photoinstabilities.
	Instead, such dynamic symmetry breaking is common to all cubic bromide and iodide (single-)perovskites thus far studied to the best of our knowledge.
	These results highlight the key role of structural chemistry in the anharmonic dynamics of halide perovskites, providing a new criterion for the design of soft optoelectronic semiconductors.
	\section*{Methods}\label{sec:Methods}
	\begin{center}
		\textbf{Electronic Structure Calculations}\\
	\end{center}
	\noindent\ac{DFT} calculations were performed with \ac{VASP} code~\cite{kresse1996} using the \ac{PAW} method.~\cite{kresse1999}
	We employed the \ac{PBE} exchange-correlation functional~\cite{perdew1996} and the \ac{TS} scheme~\cite{tkatchenko2009} -- using an iterative Hirshfeld partitioning of the charge density~\cite{bucko2013,bucko2014} -- to account for dispersive interactions.
	This setup has been shown to accurately describe the structure of \acp{HaP}.~\cite{egger2014,beck2019}
	All static calculations used an energy convergence threshold of \SI{E-6}{\electronvolt}, a plane-wave cutoff of \SI{500}{\electronvolt}, and a $\Gamma$-centered $k$-grid of $6\times6\times6$ ($6\times4\times6$) for the $Pm\bar{3}m$ ($Pnma$) structures.
	Lattice parameters were optimized by a fitting procedure using the Birch-Murnaghan equation of state~\cite{birch1947,murnaghan1944}
	The final structures used in all subsequent calculations were obtained by relaxing the ionic degrees of freedom until the maximum residual force was below \SI{E-4}{\electronvolt\per\angstrom}.
	The total and projected electronic \ac{DOS} and \ac{COHP}, were calculated by partitioning the \ac{DFT}-calculated band structure into bonding and antibonding contributions using the LOBSTER code.~\cite{maintz2016,nelson2020} 
	For this task, the \ac{DFT}-computed electronic wave functions were projected onto Slater-type orbitals (basis set name: "pbeVaspFit2015")~\cite{maintz2016} including Cs 6s, 5p and 5s, Pb 6s and 6p, and Br 4p and 4s states.
	The maximum charge spilling in this procedure was 1.3\%.
	Spin-orbit coupling was not included in our calculations, since it is currently not supported by the LOBSTER code. We emphasize that our focus is on the orbital contributions to the (anti) bonding interactions, rather than on a quantitative descriptions of the energy.\\
	\begin{center}
		\textbf{Phonon Calculations}\\
	\end{center}
	\noindent Phonon dispersions and \acp{DOS} were obtained \textit{via} the finite displacements method implemented in the phonopy package.~\cite{togo2015}
	For these calculations, we used $2\times2\times2$ supercells with 40 (160) atoms of the $Pm\bar{3}m$ ($Pnma$) \ce{CsMBr3} structures reducing $k$-space sampling accordingly.
	\ac{IR} and Raman spectra were computed with the phonopy-spectroscopy package,~\cite{skelton2017} using zone-center phonon modes, Born-effective charges and polarizabilities, calculated with \ac{DFPT}.~\cite{gajdos2006} \\
	\begin{center}
		\textbf{First-principles Molecular Dynamics}\\
	\end{center}
	\noindent\ac{DFT}-based \ac{MD} calculations were performed for $2\times2\times2$ supercells of the $Pm\bar{3}m$ structures using a Nos\'{e}-Hoover thermostat within the canonical ensemble (NVT), as implemented in \ac{VASP}.~\cite{kresse1994}
	The simulation temperature was set to $T$=\SIlist{525;570}{\kelvin} for \cpb~and \csb, respectively.
	An \SI{8}{\femto\second} timestep, reduced $k$-grid of $3\times3\times3$, and energy convergence threshold of \SI{E-5}{\electronvolt} were used for the \SI{10}{\pico\second} equilibration and \SI{115}{\pico\second} production runs.\\
\textcolor{black}{
	\begin{center}
		\textbf{Raman Spectra From Molecular Dynamics}\\
	\end{center}
	\noindent\ac{DFT}-based \ac{MD} calculations were used to compute the high-temperature Raman spectra of \cpb~and \csb. We calculated Raman intensities from the autocorrelation function of the polarisability, as detailed elsewhere.\cite{Thomas_2013}
 The polarizabilities were calculated with \ac{DFPT}~\cite{gajdos2006} on 400 evenly-spaced snapshots every 0.11\,ps for a total of 44.8\,ps. The $k$-grid employed for the \ac{DFPT} calculations was set to $4\times4\times4$ after testing convergence of the polarisability tensor for several snapshots.}\\
 
	\begin{center}
		\textbf{Octahedral Rotation Dynamics and Cross-correlations}\\
	\end{center}
	\noindent We quantified the octahedral dynamics using the rotation angles, $\phi_{\alpha}$, around a given Cartesian axis $\alpha$ (see Figure \ref{fig:VDOS}a). 
	The frequency-resolved rotational dynamics were calculated as the Fourier transform of $\phi_{\alpha}$:
	\begin{equation}\label{eq:ang-frequency}
		\Phi_{\alpha}(\omega) = \frac{1}{N_\mathrm{steps}}\int_{0}^{\infty}{\phi_{\alpha}(t) e^{-i\omega t}dt},
	\end{equation}
	where $N_{\mathrm{steps}}$ is the number of snapshots. To compute the angles we selected \num{1000} equally spaced snapshots.
	We calculated the frequency-resolved cross-correlation between octahedral rotation angles (around a Cartesian direction $\alpha$) and the displacements (along a Cartesian direction $\beta$) of the corresponding M-site, $d_{\beta}^\mathrm{M}(t)$, as:
	\begin{equation}\label{eq:Corr}
		C_{\alpha\beta}(\omega) = \frac{1}{N_\mathrm{steps}} \int_{0}^{\infty}{\frac{\langle \phi_{\alpha}(t+\delta t) \cdot d_{\beta}^\mathrm{M}(t) \rangle}{\langle \phi_{\alpha}(t) \cdot d_{\beta}^\mathrm{M}(t) \rangle}e^{-i\omega t} dt}.
	\end{equation}\\
	
	\begin{center}
		\textbf{Polycrystalline Sample Preparation}\\
	\end{center}
	\noindent \ce{CsBr} (Alfa Aesar, 99.9\%), anhydrous \ce{SrBr2} (Alfa Aesar, 99\%), \ce{Cs2CO3}, \ce{PbO}, and concentrated aqueous \ce{HBr} were purchased and used as received. Guided by the reported pseudo-binary phase diagram,~\cite{Riccardi1970} polycrystalline \csb~for X-ray powder diffraction and Raman spectroscopy was prepared by a solid-state reaction at \SI{600}{\celsius}. \ce{CsBr} (\SI{5}{\milli\mole}, \SI{1064}{\milli\gram}) and \ce{SrBr2} (\SI{5}{\milli\mole}, \SI{1237}{\milli\gram}) were ground and pressed into a \SI{5}{\milli\meter} diameter pellet, placed in an alumina crucible, and flame-sealed under $\sim$1/3 atmosphere of argon in a fused silica ampoule. The reaction yields a porous, colorless pellet which is easily separated from the crucible and ground in inert atmosphere. Polycrystalline \cpb~for X-ray powder diffraction was prepared in ambient atmosphere by precipitation from aqueous hydrobromic acid. \ce{PbO} (\SI{2}{\milli\mole}, \SI{446.4}{\milli\gram}) was dissolved in \SI{2}{\milli\liter} hot concentrated \ce{HBr} under stirring. \ce{Cs2CO3} (\SI{1}{\milli\mole}, \SI{325.8}{\milli\gram}) was added slowly resulting in an immediate bright orange precipitate. \SI{13}{\milli\liter} additional \ce{HBr} was added and the mixture left to stir. After an hour, stirring was stopped and the mixture allowed to cool to room temperature. Excess solution was decanted, and the remaining mixture was evaporated to dryness on a hotplate and ground. Phase purity of all prepared compounds was established by powder \ac{XRD}.\\
	\begin{center}
		\textbf{Single Crystal Preparation}\\
	\end{center}
	\noindent Single crystals of \csb~were grown by the Bridgman method from a stoichiometric mixture of the binary metal bromides in a \SI{10}{\milli\meter} diameter quartz ampoule. \csb~was pulled at \SI{0.5}{\milli\meter\per\hour} through an \SI{800}{\celsius} hot zone, yielding a multi-crystalline rod from which several-\unit{\milli\meter} single crystal regions could be cleaved.
	\csb~is extremely hygroscopic and all preparation and handling was performed in an inert atmosphere.\\
	The vertical Bridgman method was used to grow large, high-quality single crystals of \ce{CsPbBr3}. After synthesis and purification (see Supplementary Information for details), the ampoule was reset to the hot zone for the Bridgman Growth. The zone 1 temperature was set to \SI{650}{\celsius} with a \SI{150}{\celsius\per\hour} ramp rate, and held for \SI{12}{\hour} to ensure a full melt before sample motion occurred. The zone 2 and 3 temperatures were set to \SI{375}{\celsius}. These temperatures were held for \SI{350}{\hour} while the ampoule was moved through the furnace at a rate of \SI{0.9}{\mm\per\hour} under \SI{0.3}{\rpm} rotation. After the motion had ceased, the zone 1 temperature ramped to \SI{375}{\celsius} to make the temperature profile in the furnace uniform. The cooling program was set to slow during the phase transitions occurring near \SIlist{120;90}{\celsius}, with a \SI{10}{\celsius\per\hour} cooling rate from \SIrange{375}{175}{\celsius}, a \SI{2.5}{\celsius\per\hour} slow cooling rate from \SIrange{175}{75}{\celsius}, and a \SI{10}{\celsius\per\hour} rate to \SI{30}{\celsius}. The resulting \ce{CsPbBr3} ingot was orange-red and had large ($\geq$\SI{5}{\mm}) transparent single-crystalline domains, though the edges of some portions exhibited twinning.\\
	\begin{center}
		\textbf{X-ray Diffraction}\\
	\end{center}
	\noindent Polycrystalline samples were ground with silicon powder (as an internal standard and diluent) and packed in borosilicate glass capillaries. Powder \ac{XRD} patterns were measured in Debye$-$Scherrer geometry using a STOE Stadi P diffractometer (Mo K$_{\alpha 1}$ radiation, Ge-(111) monochromator, Mythen 1K Detector) equipped with a furnace. Data were analyzed by sequential Pawley refinement using GSAS-II.~\cite{Toby2013} \\
	
	\begin{center}
		\textbf{Infrared Reflectivity Measurements}\\
	\end{center}
	\noindent \ac{IR}-reflection spectra in the \unit{THz} range were measured as a combination of \ac{TDS} for the low-frequency end and bolometer detection for the higher frequencies.
	Bolometer spectra were measured using a Bruker 80v Fourier-transform \ac{IR} spectrometer with a globar source and a bolometer detector cooled to liquid \ce{He} temperatures. The crystals were mounted for reflection measurements and the instrument was sealed in vacuum. A gold mirror was used as reflection reference.
	\ac{TDS} was performed using a Spectra Physics Mai Tai-Empower-Spitfire Pro Ti:Sapphire regenerative amplifier. The amplifier generates \SI{35}{\femto\second} pulses centered at \SI{800}{\nm} at a repetition rate of \SI{5}{\kilo\hertz}. \unit{THz} pulses were generated by a spintronic emitter, which was composed of \SI{1.8}{\nm} of \ce{Co40Fe40B20} sandwiched between \SI{2}{\nm} of Tungsten and \SI{2}{nm} of Platinum, all supported by a quartz substrate. The \unit{THz} pulses were detected using electro-optic sampling in a (100)-\ce{ZnTe} crystal. A gold mirror was used as reflection reference. The sample crystals, \unit{THz} emitter and \unit{THz} detector were held under vacuum during the measurements.\\
	\ac{TDS} offers better signal at low frequency, while bolometer measurements have an advantage over \ac{TDS} at higher frequencies. Therefore, the spectra were combined and merged at \SI{100}{\wavenumber}. Owing to scattering losses, the absolute intensity of reflected light can not be taken quantitatively. Therefore, the spectra were scaled to the signal level at \SI{100}{\wavenumber} before merging the data. The final reflectivity spectra are given in arbitrary units. The phonon frequencies and overall spectral shape allows for fitting to the dielectric function.\\
	
	\begin{center}
		\textbf{Raman Spectroscopy}\\
	\end{center}
	\noindent 
	All the measurements were taken in a home-built back scattering Raman system.
	For all measurements, the laser was focused with a 50x objective (Zeiss, USA), and the Rayleigh scattering was then filtered with a notch filter (Ondax Inc., USA). 
	The beam was focused into a spectrometer \SI{1}{\meter} long (FHR 1000, Horiba) and then on a CCD detector.
	To get the unpolarized Raman spectrum for the single crystals (\csb~low temperatures and \cpb), two orthogonal angles were measured in parallel and cross configurations (four measurements overall). The unpolarized spectrum is a summation of all four spectra. 
	The samples were cooled below room temperature by a Janis cryostat ST-500 controlled by Lakeshore model 335 and were heated above room temperature by a closed heating system (Linkam Scientific).
	Due to the extreme sensitivity of \csb~to ambient moisture, \csb~powder was flame-sealed in a small quartz capillary for the high-temperature measurements, and a single crystal was loaded into a closed cell under an \ce{Ar} environment for the low temperatures measurements.
	\csb~low temperature measurements were taken with a \SI{2.5}{\eV} CW diode laser (\textcolor{black}{Toptica} Inc.). 
	\csb~high-temperature measurement and all the \cpb~measurements were taken with a \SI{1.57}{\eV} CW diode laser (\textcolor{black}{Toptica} Inc.).
    \textcolor{black}{We note that while Raman spectra on quartz show a contribution towards zero frequency,~\cite{bates_1972} it is narrower in frequency than what we observe. 
    Results from control experiments (see Supplementary Information) show that the main signals from quartz do not contribute to the measured Raman spectra of \csb.}
	
	\bibliographystyle{naturemag}
	\bibliography{Refs_draft}

\begin{thebibliography}{10}
\expandafter\ifx\csname url\endcsname\relax
  \def\url#1{\texttt{#1}}\fi
\expandafter\ifx\csname urlprefix\endcsname\relax\def\urlprefix{URL }\fi
\providecommand{\bibinfo}[2]{#2}
\providecommand{\eprint}[2][]{\url{#2}}

\bibitem{snaith2013}
\bibinfo{author}{Snaith, H.~J.}
\newblock \bibinfo{title}{Perovskites: {{The}} emergence of a new era for
  low-cost, high-efficiency solar cells}.
\newblock \emph{\bibinfo{journal}{J. Phys. Chem. Lett.}}
  \textbf{\bibinfo{volume}{4}}, \bibinfo{pages}{3623--3630}
  (\bibinfo{year}{2013}).

\bibitem{correa-baena2017}
\bibinfo{author}{{Correa-Baena}, J.-P.} \emph{et~al.}
\newblock \bibinfo{title}{Promises and challenges of perovskite solar cells}.
\newblock \emph{\bibinfo{journal}{Science}} \textbf{\bibinfo{volume}{358}},
  \bibinfo{pages}{739--744} (\bibinfo{year}{2017}).

\bibitem{nayak2019}
\bibinfo{author}{Nayak, P.~K.}, \bibinfo{author}{Mahesh, S.},
  \bibinfo{author}{Snaith, H.~J.} \& \bibinfo{author}{Cahen, D.}
\newblock \bibinfo{title}{Photovoltaic solar cell technologies: analysing the
  state of the art}.
\newblock \emph{\bibinfo{journal}{Nat. Rev. Mater.}}
  \textbf{\bibinfo{volume}{4}}, \bibinfo{pages}{269--285}
  (\bibinfo{year}{2019}).

\bibitem{beecher2016}
\bibinfo{author}{Beecher, A.~N.} \emph{et~al.}
\newblock \bibinfo{title}{Direct {{Observation}} of {{Dynamic Symmetry
  Breaking}} above {{Room Temperature}} in {{Methylammonium Lead Iodide
  Perovskite}}}.
\newblock \emph{\bibinfo{journal}{ACS Energy Lett.}}
  \textbf{\bibinfo{volume}{1}}, \bibinfo{pages}{880--887}
  (\bibinfo{year}{2016}).
\newblock \eprint{1606.09267}.

\bibitem{whalley2016}
\bibinfo{author}{Whalley, L.~D.}, \bibinfo{author}{Skelton, J.~M.},
  \bibinfo{author}{Frost, J.~M.} \& \bibinfo{author}{Walsh, A.}
\newblock \bibinfo{title}{Phonon anharmonicity, lifetimes, and thermal
  transport in
  {{CH}}{\textsubscript{3}}{{NH}}{\textsubscript{3}}{{PbI}}{\textsubscript{3}}
  from many-body perturbation theory}.
\newblock \emph{\bibinfo{journal}{Phys. Rev. B}} \textbf{\bibinfo{volume}{94}},
  \bibinfo{pages}{1--5} (\bibinfo{year}{2016}).

\bibitem{carignano2017}
\bibinfo{author}{Carignano, M.~A.}, \bibinfo{author}{Aravindh, S.~A.},
  \bibinfo{author}{Roqan, I.~S.}, \bibinfo{author}{Even, J.} \&
  \bibinfo{author}{Katan, C.}
\newblock \bibinfo{title}{Critical {{Fluctuations}} and {{Anharmonicity}} in
  {{Lead Iodide Perovskites}} from {{Molecular Dynamics Supercell
  Simulations}}}.
\newblock \emph{\bibinfo{journal}{J. Mater. Chem. C}}
  \textbf{\bibinfo{volume}{121}}, \bibinfo{pages}{20729--20738}
  (\bibinfo{year}{2017}).

\bibitem{brenner2016}
\bibinfo{author}{Brenner, T.~M.}, \bibinfo{author}{Egger, D.~A.},
  \bibinfo{author}{Kronik, L.}, \bibinfo{author}{Hodes, G.} \&
  \bibinfo{author}{Cahen, D.}
\newblock \bibinfo{title}{Hybrid organic\textemdash inorganic perovskites:
  low-cost semiconductors with intriguing charge-transport properties}.
\newblock \emph{\bibinfo{journal}{Nat. Rev. Mater.}}
  \textbf{\bibinfo{volume}{1}}, \bibinfo{pages}{15007} (\bibinfo{year}{2016}).
\newblock \eprint{1011.1669v3}.

\bibitem{Egger2018}
\bibinfo{author}{Egger, D.~A.} \emph{et~al.}
\newblock \bibinfo{title}{What remains unexplained about the properties of
  halide perovskites?}
\newblock \emph{\bibinfo{journal}{Adv. Mater.}} \textbf{\bibinfo{volume}{30}},
  \bibinfo{pages}{1800691} (\bibinfo{year}{2018}).

\bibitem{ceratti2018}
\bibinfo{author}{Ceratti, D.~R.} \emph{et~al.}
\newblock \bibinfo{title}{Self-{{Healing Inside APbBr}}{\textsubscript{3}}
  {{Halide Perovskite Crystals}}}.
\newblock \emph{\bibinfo{journal}{Adv. Mater.}} \textbf{\bibinfo{volume}{30}},
  \bibinfo{pages}{1706273} (\bibinfo{year}{2018}).

\bibitem{cahen2021}
\bibinfo{author}{Cahen, D.}, \bibinfo{author}{Kronik, L.} \&
  \bibinfo{author}{Hodes, G.}
\newblock \bibinfo{title}{Are {{Defects}} in {{Lead-Halide Perovskites
  Healed}}, {{Tolerated}}, or {{Both}}?}
\newblock \emph{\bibinfo{journal}{ACS Energy Lett.}}
  \bibinfo{pages}{4108--4114} (\bibinfo{year}{2021}).

\bibitem{parida2023}
\bibinfo{author}{Parida, S.} \emph{et~al.}
\newblock \bibinfo{title}{Self-{{Healing}} and -{{Repair}} of {{Nanomechanical
  Damages}} in {{Lead Halide Perovskites}}}.
\newblock \emph{\bibinfo{journal}{Adv Funct Materials}}
  \bibinfo{pages}{2304278} (\bibinfo{year}{2023}).

\bibitem{patrick2015}
\bibinfo{author}{Patrick, C.~E.}, \bibinfo{author}{Jacobsen, K.~W.} \&
  \bibinfo{author}{Thygesen, K.~S.}
\newblock \bibinfo{title}{Anharmonic stabilization and band gap renormalization
  in the perovskite {{CsSnI}}{\textsubscript{3}}}.
\newblock \emph{\bibinfo{journal}{Phys. Rev. B}} \textbf{\bibinfo{volume}{92}},
  \bibinfo{pages}{201205} (\bibinfo{year}{2015}).

\bibitem{wiktor2017}
\bibinfo{author}{Wiktor, J.}, \bibinfo{author}{Rothlisberger, U.} \&
  \bibinfo{author}{Pasquarello, A.}
\newblock \bibinfo{title}{Predictive {{Determination}} of {{Band Gaps}} of
  {{Inorganic Halide Perovskites}}}.
\newblock \emph{\bibinfo{journal}{J. Phys. Chem. Lett.}}
  \textbf{\bibinfo{volume}{8}}, \bibinfo{pages}{5507--5512}
  (\bibinfo{year}{2017}).

\bibitem{lanigan-atkins2021}
\bibinfo{author}{{Lanigan-Atkins}, T.} \emph{et~al.}
\newblock \bibinfo{title}{Two-dimensional overdamped fluctuations of the soft
  perovskite lattice in {{CsPbBr}}{\textsubscript{3}}}.
\newblock \emph{\bibinfo{journal}{Nat. Mater.}} \textbf{\bibinfo{volume}{20}},
  \bibinfo{pages}{977--983} (\bibinfo{year}{2021}).

\bibitem{seidl2023}
\bibinfo{author}{Seidl, S.~A.} \emph{et~al.}
\newblock \bibinfo{title}{Anharmonic fluctuations govern the band gap of halide
  perovskites}.
\newblock \emph{\bibinfo{journal}{Phys. Rev. Materials}}
  \textbf{\bibinfo{volume}{7}}, \bibinfo{pages}{L092401}
  (\bibinfo{year}{2023}).

\bibitem{gehrmann2019}
\bibinfo{author}{Gehrmann, C.} \& \bibinfo{author}{Egger, D.~A.}
\newblock \bibinfo{title}{Dynamic shortening of disorder potentials in
  anharmonic halide perovskites}.
\newblock \emph{\bibinfo{journal}{Nat. Commun.}} \textbf{\bibinfo{volume}{10}},
  \bibinfo{pages}{3141} (\bibinfo{year}{2019}).

\bibitem{gehrmann2022}
\bibinfo{author}{Gehrmann, C.}, \bibinfo{author}{Caicedo-D{\'a}vila, S.},
  \bibinfo{author}{Zhu, X.} \& \bibinfo{author}{Egger, D.~A.}
\newblock \bibinfo{title}{Transversal {{Halide Motion Intensifies
  Band}}-{{To}}-{{Band Transitions}} in {{Halide Perovskites}}}.
\newblock \emph{\bibinfo{journal}{Adv. Sci.}} \textbf{\bibinfo{volume}{9}},
  \bibinfo{pages}{2200706} (\bibinfo{year}{2022}).

\bibitem{wu2019}
\bibinfo{author}{Wu, B.} \emph{et~al.}
\newblock \bibinfo{title}{Indirect tail states formation by thermal-induced
  polar fluctuations in halide perovskites}.
\newblock \emph{\bibinfo{journal}{Nat. Commun.}} \textbf{\bibinfo{volume}{10}},
  \bibinfo{pages}{1--10} (\bibinfo{year}{2019}).

\bibitem{mayers2018}
\bibinfo{author}{Mayers, M.~Z.}, \bibinfo{author}{Tan, L.~Z.},
  \bibinfo{author}{Egger, D.~A.}, \bibinfo{author}{Rappe, A.~M.} \&
  \bibinfo{author}{Reichman, D.~R.}
\newblock \bibinfo{title}{How {{Lattice}} and {{Charge Fluctuations Control
  Carrier Dynamics}} in {{Halide Perovskites}}}.
\newblock \emph{\bibinfo{journal}{Nano Lett.}} \textbf{\bibinfo{volume}{18}},
  \bibinfo{pages}{8041--8046} (\bibinfo{year}{2018}).

\bibitem{lacroix2020}
\bibinfo{author}{Lacroix, A.}, \bibinfo{author}{De~Laissardi{\`e}re, G.~T.},
  \bibinfo{author}{Qu{\'e}merais, P.}, \bibinfo{author}{Julien, J.~P.} \&
  \bibinfo{author}{Mayou, D.}
\newblock \bibinfo{title}{Modeling of {{Electronic Mobilities}} in {{Halide
  Perovskites}}: {{Adiabatic Quantum Localization Scenario}}}.
\newblock \emph{\bibinfo{journal}{Phys. Rev. Lett.}}
  \textbf{\bibinfo{volume}{124}}, \bibinfo{pages}{1--6} (\bibinfo{year}{2020}).

\bibitem{iaru2021}
\bibinfo{author}{Iaru, C.~M.} \emph{et~al.}
\newblock \bibinfo{title}{Fr\"ohlich interaction dominated by a single phonon
  mode in {{CsPbBr}}{\textsubscript{3}}}.
\newblock \emph{\bibinfo{journal}{Nat. Commun.}} \textbf{\bibinfo{volume}{12}},
  \bibinfo{pages}{5844} (\bibinfo{year}{2021}).

\bibitem{schilcher2021}
\bibinfo{author}{Schilcher, M.~J.} \emph{et~al.}
\newblock \bibinfo{title}{The {{Significance}} of {{Polarons}} and {{Dynamic
  Disorder}} in {{Halide Perovskites}}}.
\newblock \emph{\bibinfo{journal}{ACS Energy Lett.}}
  \textbf{\bibinfo{volume}{6}}, \bibinfo{pages}{2162--2173}
  (\bibinfo{year}{2021}).

\bibitem{lai2022}
\bibinfo{author}{Lai, R.} \emph{et~al.}
\newblock \bibinfo{title}{Transient {{Suppression}} of {{Carrier Mobility Due}}
  to {{Hot Optical Phonons}} in {{Lead Bromide Perovskites}}}.
\newblock \emph{\bibinfo{journal}{J. Phys. Chem. Lett.}}
  \textbf{\bibinfo{volume}{13}}, \bibinfo{pages}{5488--5494}
  (\bibinfo{year}{2022}).

\bibitem{zhang2022}
\bibinfo{author}{Zhang, K.-C.}, \bibinfo{author}{Shen, C.},
  \bibinfo{author}{Zhang, H.-B.}, \bibinfo{author}{Li, Y.-F.} \&
  \bibinfo{author}{Liu, Y.}
\newblock \bibinfo{title}{Effect of quartic anharmonicity on the carrier
  transport of cubic halide perovskites {{CsSnI}}{\textsubscript{3}} and
  {{CsPbI}}{\textsubscript{3}}}.
\newblock \emph{\bibinfo{journal}{Phys. Rev. B}}
  \textbf{\bibinfo{volume}{106}}, \bibinfo{pages}{235202}
  (\bibinfo{year}{2022}).

\bibitem{schilcher2023}
\bibinfo{author}{Schilcher, M.~J.} \emph{et~al.}
\newblock \bibinfo{title}{Correlated anharmonicity and dynamic disorder control
  carrier transport in halide perovskites}.
\newblock \emph{\bibinfo{journal}{Phys. Rev. Materials}}
  \textbf{\bibinfo{volume}{7}}, \bibinfo{pages}{L081601}
  (\bibinfo{year}{2023}).

\bibitem{fabini2020}
\bibinfo{author}{Fabini, D.~H.}, \bibinfo{author}{Seshadri, R.} \&
  \bibinfo{author}{Kanatzidis, M.~G.}
\newblock \bibinfo{title}{The underappreciated lone pair in halide perovskites
  underpins their unusual properties}.
\newblock \emph{\bibinfo{journal}{MRS Bull.}} \textbf{\bibinfo{volume}{45}},
  \bibinfo{pages}{467--477} (\bibinfo{year}{2020}).

\bibitem{VanderVorst1980}
\bibinfo{author}{Van~der Vorst, C.} \& \bibinfo{author}{Maaskant, W.}
\newblock \bibinfo{title}{Stereochemically active (5s)$^2$ lone pairs in the
  structures of $\alpha$-incl and $\beta$-incl}.
\newblock \emph{\bibinfo{journal}{J. Solid State Chem.}}
  \textbf{\bibinfo{volume}{34}}, \bibinfo{pages}{301–313}
  (\bibinfo{year}{1980}).
\newblock \urlprefix\url{http://dx.doi.org/10.1016/0022-4596(80)90428-4}.

\bibitem{Maaskant1991}
\bibinfo{author}{Maaskant, W. J.~A.} \& \bibinfo{author}{Bersuker, I.~B.}
\newblock \bibinfo{title}{A combined jahn-teller and pseudo-jahn-teller effect:
  an exactly solvable model}.
\newblock \emph{\bibinfo{journal}{J. Phys. Condens. Matter}}
  \textbf{\bibinfo{volume}{3}}, \bibinfo{pages}{37–47}
  (\bibinfo{year}{1991}).
\newblock \urlprefix\url{http://dx.doi.org/10.1088/0953-8984/3/1/003}.

\bibitem{Bersuker2013}
\bibinfo{author}{Bersuker, I.~B.}
\newblock \bibinfo{title}{Pseudo-{{Jahn}}\textendash{{Teller
  Effect}}\textemdash{{A}} two-state paradigm in formation, deformation, and
  transformation of molecular systems and solids}.
\newblock \emph{\bibinfo{journal}{Chem. Rev.}} \textbf{\bibinfo{volume}{113}},
  \bibinfo{pages}{1351--1390} (\bibinfo{year}{2013}).

\bibitem{fabini2016}
\bibinfo{author}{Fabini, D.~H.} \emph{et~al.}
\newblock \bibinfo{title}{Dynamic {{Stereochemical Activity}} of the
  {{Sn}}{\textsuperscript{2+}} {{Lone Pair}} in {{Perovskite
  CsSnBr}}{\textsubscript{3}}}.
\newblock \emph{\bibinfo{journal}{J. Am. Chem. Soc.}}
  \textbf{\bibinfo{volume}{138}}, \bibinfo{pages}{11820--11832}
  (\bibinfo{year}{2016}).

\bibitem{laurita2017}
\bibinfo{author}{Laurita, G.}, \bibinfo{author}{Fabini, D.~H.},
  \bibinfo{author}{Stoumpos, C.~C.}, \bibinfo{author}{Kanatzidis, M.~G.} \&
  \bibinfo{author}{Seshadri, R.}
\newblock \bibinfo{title}{Chemical tuning of dynamic cation off-centering in
  the cubic phases of hybrid tin and lead halide perovskites}.
\newblock \emph{\bibinfo{journal}{Chem. Sci.}} \textbf{\bibinfo{volume}{8}},
  \bibinfo{pages}{5628--5635} (\bibinfo{year}{2017}).

\bibitem{gao2021}
\bibinfo{author}{Gao, L.} \emph{et~al.}
\newblock \bibinfo{title}{Metal cation s lone-pairs increase octahedral tilting
  instabilities in halide perovskites}.
\newblock \emph{\bibinfo{journal}{Mater. Adv.}} \textbf{\bibinfo{volume}{2}},
  \bibinfo{pages}{4610--4616} (\bibinfo{year}{2021}).

\bibitem{yaffe2017}
\bibinfo{author}{Yaffe, O.} \emph{et~al.}
\newblock \bibinfo{title}{Local {{Polar Fluctuations}} in {{Lead Halide
  Perovskite Crystals}}}.
\newblock \emph{\bibinfo{journal}{Phys. Rev. Lett.}}
  \textbf{\bibinfo{volume}{118}}, \bibinfo{pages}{1--6} (\bibinfo{year}{2017}).

\bibitem{huang2022}
\bibinfo{author}{Huang, X.} \emph{et~al.}
\newblock \bibinfo{title}{Understanding {{Electron}}\textendash{{Phonon
  Interactions}} in {{3D Lead Halide Perovskites}} from the {{Stereochemical
  Expression}} of 6s{\textsuperscript{2}} {{Lone Pairs}}}.
\newblock \emph{\bibinfo{journal}{J. Am. Chem. Soc.}}
  \textbf{\bibinfo{volume}{144}}, \bibinfo{pages}{12247--12260}
  (\bibinfo{year}{2022}).

\bibitem{walsh2011}
\bibinfo{author}{Walsh, A.}, \bibinfo{author}{Payne, D.~J.},
  \bibinfo{author}{Egdell, R.~G.} \& \bibinfo{author}{Watson, G.~W.}
\newblock \bibinfo{title}{Stereochemistry of post-transition metal oxides:
  revision of the classical lone pair model}.
\newblock \emph{\bibinfo{journal}{Chem. Soc. Rev.}}
  \textbf{\bibinfo{volume}{40}}, \bibinfo{pages}{4455} (\bibinfo{year}{2011}).

\bibitem{smith2015}
\bibinfo{author}{Smith, E.~H.}, \bibinfo{author}{Benedek, N.~A.} \&
  \bibinfo{author}{Fennie, C.~J.}
\newblock \bibinfo{title}{Interplay of {{Octahedral Rotations}} and {{Lone Pair
  Ferroelectricity}} in {{CsPbF}}{\textsubscript{3}}}.
\newblock \emph{\bibinfo{journal}{Inorg. Chem.}} \textbf{\bibinfo{volume}{54}},
  \bibinfo{pages}{8536--8543} (\bibinfo{year}{2015}).

\bibitem{radha2018}
\bibinfo{author}{Radha, S.~K.}, \bibinfo{author}{Bhandari, C.} \&
  \bibinfo{author}{Lambrecht, W.~R.}
\newblock \bibinfo{title}{Distortion modes in halide perovskites: {{To}} twist
  or to stretch, a matter of tolerance and lone pairs}.
\newblock \emph{\bibinfo{journal}{Phys. Rev. Materials}}
  \textbf{\bibinfo{volume}{2}}, \bibinfo{pages}{1--11} (\bibinfo{year}{2018}).

\bibitem{du2014}
\bibinfo{author}{Du, M.~H.}
\newblock \bibinfo{title}{Efficient carrier transport in halide perovskites:
  theoretical perspectives}.
\newblock \emph{\bibinfo{journal}{J. Mater. Chem. A}}
  \textbf{\bibinfo{volume}{2}}, \bibinfo{pages}{9091--9098}
  (\bibinfo{year}{2014}).

\bibitem{herz2018}
\bibinfo{author}{Herz, L.~M.}
\newblock \bibinfo{title}{How {{Lattice Dynamics Moderate}} the {{Electronic
  Properties}} of {{Metal-Halide Perovskites}}}.
\newblock \emph{\bibinfo{journal}{J. Phys. Chem. Lett.}}
  \textbf{\bibinfo{volume}{9}}, \bibinfo{pages}{6853--6863}
  (\bibinfo{year}{2018}).

\bibitem{goldschmidt1926}
\bibinfo{author}{Goldschmidt, V.~M.}
\newblock \bibinfo{title}{{Die Gesetze der Krystallochemie}}.
\newblock \emph{\bibinfo{journal}{Naturwissenschaften}}
  \textbf{\bibinfo{volume}{14}}, \bibinfo{pages}{477--485}
  (\bibinfo{year}{1926}).

\bibitem{Cohen2022}
\bibinfo{author}{Cohen, A.} \emph{et~al.}
\newblock \bibinfo{title}{Diverging expressions of anharmonicity in halide
  perovskites}.
\newblock \emph{\bibinfo{journal}{Adv. Mater.}} \textbf{\bibinfo{volume}{34}},
  \bibinfo{pages}{2107932} (\bibinfo{year}{2022}).

\bibitem{ferreira2020}
\bibinfo{author}{Ferreira, A.~C.} \emph{et~al.}
\newblock \bibinfo{title}{Direct evidence of weakly dispersed and strongly
  anharmonic optical phonons in hybrid perovskites}.
\newblock \emph{\bibinfo{journal}{Commun. Phys.}} \textbf{\bibinfo{volume}{3}},
  \bibinfo{pages}{48} (\bibinfo{year}{2020}).

\bibitem{sharma2020}
\bibinfo{author}{Sharma, R.} \emph{et~al.}
\newblock \bibinfo{title}{Elucidating the atomistic origin of anharmonicity in
  tetragonal
  {{CH}}{\textsubscript{3}}{{NH}}{\textsubscript{3}}{{PbI}}{\textsubscript{3}}
  with {{Raman}} scattering}.
\newblock \emph{\bibinfo{journal}{Phys. Rev. Materials}}
  \textbf{\bibinfo{volume}{4}}, \bibinfo{pages}{092401} (\bibinfo{year}{2020}).

\bibitem{menahem2023}
\bibinfo{author}{Menahem, M.} \emph{et~al.}
\newblock \bibinfo{title}{Disorder origin of {{Raman}} scattering in perovskite
  single crystals}.
\newblock \emph{\bibinfo{journal}{Phys. Rev. Materials}}
  \textbf{\bibinfo{volume}{7}}, \bibinfo{pages}{044602} (\bibinfo{year}{2023}).

\bibitem{goesten2018}
\bibinfo{author}{Goesten, M.~G.} \& \bibinfo{author}{Hoffmann, R.}
\newblock \bibinfo{title}{Mirrors of {{Bonding}} in {{Metal Halide
  Perovskites}}}.
\newblock \emph{\bibinfo{journal}{J. Am. Chem. Soc.}}
  \textbf{\bibinfo{volume}{140}}, \bibinfo{pages}{12996--13010}
  (\bibinfo{year}{2018}).

\bibitem{straus2022}
\bibinfo{author}{Straus, D.~B.} \& \bibinfo{author}{Cava, R.~J.}
\newblock \bibinfo{title}{Tuning the {{Band Gap}} in the {{Halide Perovskite
  CsPbBr}}{\textsubscript{3}} through {{Sr Substitution}}}.
\newblock \emph{\bibinfo{journal}{ACS Appl. Mater. Interfaces}}
  (\bibinfo{year}{2022}).

\bibitem{du2010}
\bibinfo{author}{Du, M.-H.} \& \bibinfo{author}{Singh, D.~J.}
\newblock \bibinfo{title}{Enhanced {{Born}} charge and proximity to
  ferroelectricity in thallium halides}.
\newblock \emph{\bibinfo{journal}{Phys. Rev. B}} \textbf{\bibinfo{volume}{81}},
  \bibinfo{pages}{144114} (\bibinfo{year}{2010}).

\bibitem{sun2017}
\bibinfo{author}{Sun, J.} \& \bibinfo{author}{Singh, D.~J.}
\newblock \bibinfo{title}{Electronic {{Properties}}, {{Screening}}, and
  {{Efficient Carrier Transport}} in {{NaSbS}}{\textsubscript{2}}}.
\newblock \emph{\bibinfo{journal}{Phys. Rev. Applied}}
  \textbf{\bibinfo{volume}{7}}, \bibinfo{pages}{024015} (\bibinfo{year}{2017}).

\bibitem{ran2018}
\bibinfo{author}{Ran, Z.} \emph{et~al.}
\newblock \bibinfo{title}{Bismuth and antimony-based oxyhalides and
  chalcohalides as potential optoelectronic materials}.
\newblock \emph{\bibinfo{journal}{npj Comput Mater}}
  \textbf{\bibinfo{volume}{4}}, \bibinfo{pages}{14} (\bibinfo{year}{2018}).

\bibitem{kang2018}
\bibinfo{author}{Kang, B.} \& \bibinfo{author}{Biswas, K.}
\newblock \bibinfo{title}{Exploring {{Polaronic}}, {{Excitonic Structures}} and
  {{Luminescence}} in
  {{Cs}}{\textsubscript{4}}{{PbBr}}{\textsubscript{6}}/{{CsPbBr}}{\textsubscript{3}}}.
\newblock \emph{\bibinfo{journal}{J. Phys. Chem. Lett.}}
  \textbf{\bibinfo{volume}{9}}, \bibinfo{pages}{830--836}
  (\bibinfo{year}{2018}).

\bibitem{rodova2003}
\bibinfo{author}{Rodov{\'a}, M.}, \bibinfo{author}{Bro{\v z}ek, J.} \&
  \bibinfo{author}{Nitsch, K.}
\newblock \bibinfo{title}{Phase transitions in ternary caesium lead bromide}.
\newblock \emph{\bibinfo{journal}{J. Therm. Anal.}}
  \textbf{\bibinfo{volume}{71}}, \bibinfo{pages}{667--673}
  (\bibinfo{year}{2003}).

\bibitem{stoumpos2013}
\bibinfo{author}{Stoumpos, C.~C.} \emph{et~al.}
\newblock \bibinfo{title}{Crystal {{Growth}} of the {{Perovskite Semiconductor
  CsPbBr}}{\textsubscript{3}}: {{A New Material}} for {{High-Energy Radiation
  Detection}}}.
\newblock \emph{\bibinfo{journal}{Cryst. Growth Des.}}
  \textbf{\bibinfo{volume}{13}}, \bibinfo{pages}{2722--2727}
  (\bibinfo{year}{2013}).

\bibitem{loyd2018}
\bibinfo{author}{Loyd, M.} \emph{et~al.}
\newblock \bibinfo{title}{Crystal structure and thermal expansion of
  {{CsCaI}}{\textsubscript{3}}:{{Eu}} and {{CsSrBr}}{\textsubscript{3}}:{{Eu}}
  scintillators}.
\newblock \emph{\bibinfo{journal}{J. Cryst. Growth}}
  \textbf{\bibinfo{volume}{481}}, \bibinfo{pages}{35--39}
  (\bibinfo{year}{2018}).

\bibitem{fabini2016a}
\bibinfo{author}{Fabini, D.~H.} \emph{et~al.}
\newblock \bibinfo{title}{Reentrant {{Structural}} and {{Optical Properties}}
  and {{Large Positive Thermal Expansion}} in {{Perovskite Formamidinium Lead
  Iodide}}}.
\newblock \emph{\bibinfo{journal}{Angew. Chem. Int. Ed.}}
  \textbf{\bibinfo{volume}{55}}, \bibinfo{pages}{15392--15396}
  (\bibinfo{year}{2016}).

\bibitem{schueller2018}
\bibinfo{author}{Schueller, E.~C.} \emph{et~al.}
\newblock \bibinfo{title}{Crystal {{Structure Evolution}} and {{Notable Thermal
  Expansion}} in {{Hybrid Perovskites Formamidinium Tin Iodide}} and
  {{Formamidinium Lead Bromide}}}.
\newblock \emph{\bibinfo{journal}{Inorg. Chem.}} \textbf{\bibinfo{volume}{57}},
  \bibinfo{pages}{695--701} (\bibinfo{year}{2018}).

\bibitem{sendner2016}
\bibinfo{author}{Sendner, M.} \emph{et~al.}
\newblock \bibinfo{title}{Optical phonons in methylammonium lead halide
  perovskites and implications for charge transport}.
\newblock \emph{\bibinfo{journal}{Mater. Horiz.}} \textbf{\bibinfo{volume}{3}},
  \bibinfo{pages}{613--620} (\bibinfo{year}{2016}).

\bibitem{armstrong1989}
\bibinfo{author}{Armstrong, R.~L.}
\newblock \bibinfo{title}{Displacive order\textemdash disorder crossover in
  perovskite and antifluorite crystals undergoing rotational phase
  transitions}.
\newblock \emph{\bibinfo{journal}{Prog. Nucl. Magn. Reson. Spectrosc.}}
  \textbf{\bibinfo{volume}{21}}, \bibinfo{pages}{151--173}
  (\bibinfo{year}{1989}).

\bibitem{woodward1997}
\bibinfo{author}{Woodward, P.~M.}
\newblock \bibinfo{title}{Octahedral {{Tilting}} in {{Perovskites}}. {{I}}.
  {{Geometrical Considerations}}}.
\newblock \emph{\bibinfo{journal}{Acta Crystallogr B}}
  \textbf{\bibinfo{volume}{53}}, \bibinfo{pages}{32--43}
  (\bibinfo{year}{1997}).

\bibitem{yang2017a}
\bibinfo{author}{Yang, R.~X.}, \bibinfo{author}{Skelton, J.~M.},
  \bibinfo{author}{Da~Silva, E.~L.}, \bibinfo{author}{Frost, J.~M.} \&
  \bibinfo{author}{Walsh, A.}
\newblock \bibinfo{title}{Spontaneous octahedral tilting in the cubic inorganic
  cesium halide perovskites {{CsSnX}}{\textsubscript{3}} and
  {{CsPbX}}{\textsubscript{3}} ({{X}} = {{F}}, {{Cl}}, {{Br}}, {{I}})}.
\newblock \emph{\bibinfo{journal}{J. Phys. Chem. Lett.}}
  \textbf{\bibinfo{volume}{8}}, \bibinfo{pages}{4720--4726}
  (\bibinfo{year}{2017}).

\bibitem{yang2020}
\bibinfo{author}{Yang, R.~X.}, \bibinfo{author}{Skelton, J.~M.},
  \bibinfo{author}{{da Silva}, E.~L.}, \bibinfo{author}{Frost, J.~M.} \&
  \bibinfo{author}{Walsh, A.}
\newblock \bibinfo{title}{Assessment of dynamic structural instabilities across
  24 cubic inorganic halide perovskites}.
\newblock \emph{\bibinfo{journal}{J. Chem. Phys.}}
  \textbf{\bibinfo{volume}{152}}, \bibinfo{pages}{024703}
  (\bibinfo{year}{2020}).

\bibitem{klarbring2019}
\bibinfo{author}{Klarbring, J.}
\newblock \bibinfo{title}{Low-energy paths for octahedral tilting in inorganic
  halide perovskites}.
\newblock \emph{\bibinfo{journal}{Phys. Rev. B}} \textbf{\bibinfo{volume}{99}},
  \bibinfo{pages}{1--7} (\bibinfo{year}{2019}).
\newblock \eprint{1802.09632}.

\bibitem{bechtel2019}
\bibinfo{author}{Bechtel, J.~S.}, \bibinfo{author}{Thomas, J.~C.} \&
  \bibinfo{author}{Van Der~Ven, A.}
\newblock \bibinfo{title}{Finite-temperature simulation of anharmonicity and
  octahedral tilting transitions in halide perovskites}.
\newblock \emph{\bibinfo{journal}{Phys. Rev. Materials}}
  \textbf{\bibinfo{volume}{3}}, \bibinfo{pages}{113605} (\bibinfo{year}{2019}).

\bibitem{zhu2022}
\bibinfo{author}{Zhu, X.}, \bibinfo{author}{{Caicedo-D{\'a}vila}, S.},
  \bibinfo{author}{Gehrmann, C.} \& \bibinfo{author}{Egger, D.~A.}
\newblock \bibinfo{title}{Probing the {{Disorder Inside}} the {{Cubic Unit
  Cell}} of {{Halide Perovskites}} from {{First-Principles}}}.
\newblock \emph{\bibinfo{journal}{ACS Appl. Mater. Interfaces}}
  \textbf{\bibinfo{volume}{14}}, \bibinfo{pages}{22973--22981}
  (\bibinfo{year}{2022}).

\bibitem{wiktor2023}
\bibinfo{author}{Wiktor, J.}, \bibinfo{author}{Fransson, E.},
  \bibinfo{author}{Kubicki, D.} \& \bibinfo{author}{Erhart, P.}
\newblock \bibinfo{title}{Quantifying {{Dynamic Tilting}} in {{Halide
  Perovskites}}: {{Chemical Trends}} and {{Local Correlations}}}
  (\bibinfo{year}{2023}).
\newblock \urlprefix\url{http://arxiv.org/abs/2304.07402}.
\newblock \eprint{2304.07402}.

\bibitem{baldwin2023}
\bibinfo{author}{Baldwin, W.~J.} \emph{et~al.}
\newblock \bibinfo{title}{Dynamic {{Local Structure}} in {{Caesium Lead
  Iodide}}: {{Spatial Correlation}} and {{Transient Domains}}}.
\newblock \emph{\bibinfo{journal}{Small}} \bibinfo{pages}{2303565}
  (\bibinfo{year}{2023}).

\bibitem{Reuveni2023}
\bibinfo{author}{Reuveni, G.} \emph{et~al.}
\newblock \bibinfo{title}{Static and dynamic disorder in formamidinium lead
  bromide single crystals}.
\newblock \emph{\bibinfo{journal}{J. Phys. Chem. Lett.}}
  \textbf{\bibinfo{volume}{14}}, \bibinfo{pages}{1288--1293}
  (\bibinfo{year}{2023}).

\bibitem{Kaltzoglou2016}
\bibinfo{author}{Kaltzoglou, A.} \emph{et~al.}
\newblock \bibinfo{title}{Optical-vibrational properties of the
  {{Cs}}{\textsubscript{2}}{{SnX}}{\textsubscript{6}} ({{X}} = {{Cl}}, {{Br}},
  {{I}}) defect perovskites and hole-transport efficiency in dye-sensitized
  solar cells}.
\newblock \emph{\bibinfo{journal}{J. Phys. Chem. C}}
  \textbf{\bibinfo{volume}{120}}, \bibinfo{pages}{11777--11785}
  (\bibinfo{year}{2016}).

\bibitem{Thiele1987}
\bibinfo{author}{Thiele, G.}, \bibinfo{author}{Rotter, H.~W.} \&
  \bibinfo{author}{Schmidt, K.~D.}
\newblock \bibinfo{title}{Kristallstrukturen und phasentransformationen von
  caesiumtrihalogenogermanaten(ii) csgex3 (x = cl, br, i)}.
\newblock \emph{\bibinfo{journal}{Z. Anorg. Allg. Chem.}}
  \textbf{\bibinfo{volume}{545}}, \bibinfo{pages}{148–156}
  (\bibinfo{year}{1987}).
\newblock \urlprefix\url{http://dx.doi.org/10.1002/zaac.19875450217}.

\bibitem{kresse1996}
\bibinfo{author}{Kresse, G.} \& \bibinfo{author}{Furthm{\"u}ller, J.}
\newblock \bibinfo{title}{Efficient iterative schemes for ab initio
  total-energy calculations using a plane-wave basis set}.
\newblock \emph{\bibinfo{journal}{Phys. Rev. B}} \textbf{\bibinfo{volume}{54}},
  \bibinfo{pages}{11169--11186} (\bibinfo{year}{1996}).

\bibitem{kresse1999}
\bibinfo{author}{Kresse, G.} \& \bibinfo{author}{Joubert, D.}
\newblock \bibinfo{title}{From ultrasoft pseudopotentials to the projector
  augmented-wave method}.
\newblock \emph{\bibinfo{journal}{Phys. Rev. B}} \textbf{\bibinfo{volume}{59}},
  \bibinfo{pages}{1758--1775} (\bibinfo{year}{1999}).

\bibitem{perdew1996}
\bibinfo{author}{Perdew, J.~P.}, \bibinfo{author}{Burke, K.} \&
  \bibinfo{author}{Ernzerhof, M.}
\newblock \bibinfo{title}{Generalized gradient approximation made simple}.
\newblock \emph{\bibinfo{journal}{Phys. Rev. Lett.}}
  \textbf{\bibinfo{volume}{77}}, \bibinfo{pages}{3865--3868}
  (\bibinfo{year}{1996}).

\bibitem{tkatchenko2009}
\bibinfo{author}{Tkatchenko, A.} \& \bibinfo{author}{Scheffler, M.}
\newblock \bibinfo{title}{Accurate {{Molecular Van Der Waals Interactions}}
  from {{Ground-State Electron Density}} and {{Free-Atom Reference Data}}}.
\newblock \emph{\bibinfo{journal}{Phys. Rev. Lett.}}
  \textbf{\bibinfo{volume}{102}}, \bibinfo{pages}{073005}
  (\bibinfo{year}{2009}).

\bibitem{bucko2013}
\bibinfo{author}{Bu{\v c}ko, T.}, \bibinfo{author}{Leb{\`e}gue, S.},
  \bibinfo{author}{Hafner, J.} \& \bibinfo{author}{{\'A}ngy{\'a}n, J.~G.}
\newblock \bibinfo{title}{Improved density dependent correction for the
  description of {{London}} dispersion forces}.
\newblock \emph{\bibinfo{journal}{J. Chem. Theory Comput.}}
  \textbf{\bibinfo{volume}{9}}, \bibinfo{pages}{4293--4299}
  (\bibinfo{year}{2013}).

\bibitem{bucko2014}
\bibinfo{author}{Bu{\v c}ko, T.}, \bibinfo{author}{Leb{\`e}gue, S.},
  \bibinfo{author}{{\'A}ngy{\'a}n, J.~G.} \& \bibinfo{author}{Hafner, J.}
\newblock \bibinfo{title}{Extending the applicability of the
  {{Tkatchenko-Scheffler}} dispersion correction via iterative {{Hirshfeld}}
  partitioning}.
\newblock \emph{\bibinfo{journal}{J. Chem. Phys.}}
  \textbf{\bibinfo{volume}{141}}, \bibinfo{pages}{034114}
  (\bibinfo{year}{2014}).

\bibitem{egger2014}
\bibinfo{author}{Egger, D.~A.} \& \bibinfo{author}{Kronik, L.}
\newblock \bibinfo{title}{Role of dispersive interactions in determining
  structural properties of organic-inorganic halide perovskites: {{Insights}}
  from first-principles calculations}.
\newblock \emph{\bibinfo{journal}{J. Phys. Chem. Lett.}}
  \textbf{\bibinfo{volume}{5}}, \bibinfo{pages}{2728--2733}
  (\bibinfo{year}{2014}).

\bibitem{beck2019}
\bibinfo{author}{Beck, H.}, \bibinfo{author}{Gehrmann, C.} \&
  \bibinfo{author}{Egger, D.~A.}
\newblock \bibinfo{title}{Structure and binding in halide perovskites:
  {{Analysis}} of static and dynamic effects from dispersion-corrected density
  functional theory}.
\newblock \emph{\bibinfo{journal}{APL Mater.}} \textbf{\bibinfo{volume}{7}},
  \bibinfo{pages}{021108} (\bibinfo{year}{2019}).

\bibitem{birch1947}
\bibinfo{author}{Birch, F.}
\newblock \bibinfo{title}{Finite {{Elastic Strain}} of {{Cubic Crystals}}}.
\newblock \emph{\bibinfo{journal}{Phys. Rev.}} \textbf{\bibinfo{volume}{71}},
  \bibinfo{pages}{809--824} (\bibinfo{year}{1947}).

\bibitem{murnaghan1944}
\bibinfo{author}{Murnaghan, F.~D.}
\newblock \bibinfo{title}{The {{Compressibility}} of {{Media}} under {{Extreme
  Pressures}}}.
\newblock \emph{\bibinfo{journal}{Proc. Natl. Acad. Sci. U.S.A.}}
  \textbf{\bibinfo{volume}{30}}, \bibinfo{pages}{244--247}
  (\bibinfo{year}{1944}).

\bibitem{maintz2016}
\bibinfo{author}{Maintz, S.}, \bibinfo{author}{Deringer, V.~L.},
  \bibinfo{author}{Tchougr{\'e}eff, A.~L.} \& \bibinfo{author}{Dronskowski, R.}
\newblock \bibinfo{title}{{{LOBSTER}}: {{A}} tool to extract chemical bonding
  from plane-wave based {{DFT}}}.
\newblock \emph{\bibinfo{journal}{J. Comput. Chem.}}
  \textbf{\bibinfo{volume}{37}}, \bibinfo{pages}{1030--1035}
  (\bibinfo{year}{2016}).

\bibitem{nelson2020}
\bibinfo{author}{Nelson, R.} \emph{et~al.}
\newblock \bibinfo{title}{{{{\textsc{LOBSTER}}}} : {{Local}} orbital
  projections, atomic charges, and chemical-bonding analysis from
  {\textsc{projector-augmented-wave-based}} density-functional theory}.
\newblock \emph{\bibinfo{journal}{J Comput Chem}}
  \textbf{\bibinfo{volume}{41}}, \bibinfo{pages}{1931--1940}
  (\bibinfo{year}{2020}).

\bibitem{togo2015}
\bibinfo{author}{Togo, A.} \& \bibinfo{author}{Tanaka, I.}
\newblock \bibinfo{title}{First principles phonon calculations in materials
  science}.
\newblock \emph{\bibinfo{journal}{Scripta Materialia}}
  \textbf{\bibinfo{volume}{108}}, \bibinfo{pages}{1--5} (\bibinfo{year}{2015}).

\bibitem{skelton2017}
\bibinfo{author}{Skelton, J.~M.} \emph{et~al.}
\newblock \bibinfo{title}{Lattice dynamics of the tin sulphides
  {{SnS}}{\textsubscript{2}}, {{SnS}} and
  {{Sn}}{\textsubscript{2}}{{S}}{\textsubscript{3}}: vibrational spectra and
  thermal transport}.
\newblock \emph{\bibinfo{journal}{Phys. Chem. Chem. Phys.}}
  \textbf{\bibinfo{volume}{19}}, \bibinfo{pages}{12452--12465}
  (\bibinfo{year}{2017}).

\bibitem{gajdos2006}
\bibinfo{author}{Gajdo{\v s}, M.}, \bibinfo{author}{Hummer, K.},
  \bibinfo{author}{Kresse, G.}, \bibinfo{author}{Furthm{\"u}ller, J.} \&
  \bibinfo{author}{Bechstedt, F.}
\newblock \bibinfo{title}{Linear optical properties in the projector-augmented
  wave methodology}.
\newblock \emph{\bibinfo{journal}{Phys. Rev. B}} \textbf{\bibinfo{volume}{73}},
  \bibinfo{pages}{1--9} (\bibinfo{year}{2006}).

\bibitem{kresse1994}
\bibinfo{author}{Kresse, G.} \& \bibinfo{author}{Hafner, J.}
\newblock \bibinfo{title}{Ab initio molecular-dynamics simulation of the
  liquid-metalamorphous- semiconductor transition in germanium}.
\newblock \emph{\bibinfo{journal}{Phys. Rev. B}} \textbf{\bibinfo{volume}{49}},
  \bibinfo{pages}{14251--14269} (\bibinfo{year}{1994}).

\bibitem{Thomas_2013}
\bibinfo{author}{Thomas, M.}, \bibinfo{author}{Brehm, M.},
  \bibinfo{author}{Fligg, R.}, \bibinfo{author}{Vöhringer, P.} \&
  \bibinfo{author}{Kirchner, B.}
\newblock \bibinfo{title}{Computing vibrational spectra from ab initio
  molecular dynamics}.
\newblock \emph{\bibinfo{journal}{Phys. Chem. Chem. Phys.}}
  \textbf{\bibinfo{volume}{15}}, \bibinfo{pages}{6608--6622}
  (\bibinfo{year}{2013}).
\newblock \urlprefix\url{http://dx.doi.org/10.1039/C3CP44302G}.

\bibitem{Riccardi1970}
\bibinfo{author}{Riccardi, R.}, \bibinfo{author}{Sinistri, C.},
  \bibinfo{author}{Campari, G.~Y.} \& \bibinfo{author}{Magistris, A.}
\newblock \bibinfo{title}{Binary systems formed by alkali bromides with barium
  or strontium bromide}.
\newblock \emph{\bibinfo{journal}{Z. Naturforsch. A}}
  \textbf{\bibinfo{volume}{25}}, \bibinfo{pages}{781--785}
  (\bibinfo{year}{1970}).

\bibitem{Toby2013}
\bibinfo{author}{Toby, B.~H.} \& \bibinfo{author}{Dreele, R. B.~V.}
\newblock \bibinfo{title}{{{{\emph{GSAS-II}}}}: the genesis of a modern
  open-source all purpose crystallography software package}.
\newblock \emph{\bibinfo{journal}{J. Appl. Crystallogr.}}
  \textbf{\bibinfo{volume}{46}}, \bibinfo{pages}{544--549}
  (\bibinfo{year}{2013}).

\bibitem{bates_1972}
\bibinfo{author}{Bates, J.~B.} \& \bibinfo{author}{Quist, A.~S.}
\newblock \bibinfo{title}{{Polarized Raman Spectra of $\beta$-Quartz}}.
\newblock \emph{\bibinfo{journal}{J. Chem. Phys.}}
  \textbf{\bibinfo{volume}{56}}, \bibinfo{pages}{1528--1533}
  (\bibinfo{year}{1972}).

\end{thebibliography}
	
	\begin{center}
		\textbf{Acknowledgements}\\
	\end{center}
	\noindent Funding provided by the Alexander von Humboldt-Foundation in the framework of the Sofja Kovalevskaja Award, endowed by the German Federal Ministry of Education and Research, by the Deutsche Forschungsgemeinschaft (DFG, German Research Foundation) \textit{via} Germany's Excellence Strategy - EXC 2089/1-390776260, and by TU Munich - IAS, funded by the German Excellence Initiative and the European Union Seventh Framework Programme under Grant Agreement No. 291763, are gratefully acknowledged. Funding was provided by the Engineering and Physical Sciences Research Council (EPSRC), UK.
	The Gauss Centre for Supercomputing e.V. is acknowledged for providing computing time through the John von Neumann Institute for Computing on the GCS Supercomputer JUWELS at J\"ulich Supercomputing Centre. D.H.F. gratefully acknowledges financial support from the Alexander von Humboldt Foundation and the Max Planck Society. D.H.F. thanks Maximilian A. Plass for assistance with flame-sealing the Raman capillaries.
	K.M.M. and M.V.K. acknowledge financial support by ETH Zürich through the ETH+ Project SynMatLab ``Laboratory for Multiscale Materials Synthesis.''\\
	\begin{center}
		\textbf{Author contributions}\\
	\end{center}
	\noindent S.C.-D. performed the theoretical calculations, analyzed the data and wrote the initial manuscript under the supervision of D.A.E \textcolor{black}{and with additional support by M.G.}
	A.C. performed Raman measurements and analyzed data under the supervision of O.Y.
	S.M. performed IR measurements and analyzed data under the supervision of L.M.H.
	D.H.F. prepared polycrystalline samples, performed XRD measurements, and analyzed data.
	M.I. prepared the \csb~single crystals.
	K.M.M. prepared the \cpb~single crystals under the supervision of M.V.K.
	D.H.F. and D.A.E. conceived and supervised the project.
\end{document}